\documentclass[10pt,british]{article}
\usepackage[a4paper]{geometry}
\usepackage{babel}
\usepackage{amsmath}
\usepackage{amssymb}
\usepackage{graphicx}
\usepackage[unicode=true,
 bookmarks=false,
 breaklinks=false,pdfborder={0 0 1},backref=false,colorlinks=false]
 {hyperref}
\hypersetup{
 colorlinks,citecolor=black,filecolor=black,linkcolor=black,urlcolor=black}

\makeatletter

\providecommand{\tabularnewline}{\\}

\usepackage{amsfonts}
\usepackage[rightcaption]{sidecap}

\usepackage{subcaption}

\usepackage{bbold}

\makeatother

\begin{document}
\title{Localised pair formation in bosonic flat-band Hubbard models}
\author{Jacob Fronk\\jf@math.ku.dk\\
  https://orcid.org/0000-0002-1441-7187\\
  ~\\
  Andreas Mielke\\mielke@tphys.uni-heidelberg.de\\
  https://orcid.org/0000-0002-1008-5891\\
  ~\\
Institut f\"{u}r Theoretische Physik\\
University of Heidelberg \\
Philosophenweg 19\\
D-69121 Heidelberg, Germany}
\maketitle

\begin{abstract}
Using a generalised version of Gershgorin's circle theorem, rigorous
boundaries on the energies of the lowest states of a broad class of
line graphs above a critical filling are derived for hardcore bosonic
systems. Also a lower boundary on the energy gap towards the next
lowest states is established. Additionally, it is shown that the corresponding
eigenstates are dominated by a subspace spanned by states containing
a compactly localised pair and a lower boundary for the overlap is
derived as well. Overall, this strongly suggests localised pair formationin the
ground states of the broad class of line graphs and rigorously
proves it for some of the graphs in it, including the inhomogeneous
chequerboard chain as well as two novel examples of regular two dimensional
graphs.
\end{abstract}

\section{Introduction}

\subsection{Bosons in flat band systems}

Due to the recent progress of both theoretical \cite{Suto03}-\cite{MielkeHCB}
and experimental \cite{bosonic theory 2,bosonic theory 5,experiments1,experiments2}
nature, strongly correlated bosons on lattices have been gathering
a lot of attention over the last few years. This has lead to improved
understanding of several phenomena, including the Bose condensation
and more recently the repulsive bosonic pair formation \cite{pairformation1}-\cite{degeneracy}.
These papers investigate pair formation in very special classes of
flat band systems.

A prototypical class of lattices with a flat band are line graphs
with the kagome lattice as a prominent example \cite{Mielke91a}.
Flat band models have been studied theoretically for more than 30
years to investigate phenomena in strongly correlated systems like
ferrimagnetism \cite{LiebFerro}, ferromagnetism \cite{Mielke91a,saturated,decorated,MT93},
or macroscopic magnetization jumps \cite{Schulenburg2002}. Since
approximately ten years, they are studied also experimentally, mainly
with the help of optical lattices, see e.g. \cite{StamperKurn2012}.
And recently, flat bands are also studied in twisted bylayer graphene
at some magic angles, see e.g. \cite{TBL1}.

Flat bands are convenient to study strong correlation phenomena. Strong
correlations occur if the interaction is large compared to the band
width. In a flat band system, the latter vanishes. Therefore, an arbitrary
small interaction is already sufficient to produce strong correlation
effects. That explains the huge interest in flat band systems. On
the other hand, all typical approximation methods that are based on
perturbative approaches or mean field approaches will fail in flat
bands due to the high degeneracy in the single particle spectrum.
Therefore, many studies of flat band systems use mathematically rigorous
methods.

Interestingly, the majority of rigorous results in flat band systems
stems from fermionic systems or spin systems, whereas for bosonic
systems less is known. One rigorous result for bosons in a flat band
is the formation of a Wigner crystal at a critical density \cite{homogeneous},
which is valid for a large class of two-dimensional line graphs. Even
below this critical density a full classification of all ground states
is possible and one obtains a system with a residual entropy.

The above mentioned papers on bosonic pair formation in flat bands
\cite{pairformation1}-\cite{degeneracy} focus on specific lattices
and are to some degree based on numerical analysis or approximations
without rigid error bounds. A rigorous analytical proof on the other
hand, even in specific systems, has proven to be difficult. One important step towards this goal was achieved by Mielke \cite{MielkeHCB}, who
proved pair formation for hard core bosons if one adds one particle
to the system at the critical density of the Wigner crystal. But even
there, the class of lattices is restricted to line graphs of graphs
consisting of elementary cycles of length $4$ and with an additional
condition on the hopping matrix elements. The hopping $t'$ between
the elementary cycles of length $4$ must be sufficiently smaller
than the hopping $t$ on those cycles. We will provide a complete
explanation of these hopping terms and of the construction of the
lattices below.

\subsection{Purpose of the present paper}

This paper substantially generalises the class of graphs considered
in \cite{MielkeHCB} to include, among other things, two dimensional
graphs with doubly periodic boundary conditions, and fills a gap in
Mielke's proofs by showing that the preconditions of his second theorem,
which concerns the overlap of ground states with states containing
a localised pair, are actually fulfilled under the conditions of the
first theorem, thereby completing a first rigorous proof for the existence
of localised pair formation in the ground state of certain systems.
In addition to the chequerboard chain, where only the aforementioned
gap was missing for such a proof, the systems with completely provable
localised pair formation also include some novel two dimensional graphs,
of which two are explicitly constructed.

Furthermore, several improvements on a qualitative as well as a quantitative
level have been achieved. By introducing an asymmetrical norm and
a simplified partitioning of the Fock space, we were able to increase
the regime in which the theorem can be applied significantly, proved
that both the energy of the lowest state and the gap to the next highest
states are constant to first order in the secondary hopping parameter
$t'$ and were able to derive a concrete lower boundary for the overlap
of the ground states with those of the uncoupled system (i.e. $t'=0$.)
While the complete proofs of localised pair formation make use of
a specific local symmetry and some form of a global translational
or rotational invariance, the main result of this paper, which already
strongly suggests the existence of localised pair formation, does
not require any kind of global symmetry in the class of graphs to
which it applies.

\subsection{The Hubbard model}

The Hamiltonian of the bosonic Hubbard model is given by 
\begin{equation}
H=\sum_{\{i,j\}\in E}t_{ij}b_{j}^{\dagger}b_{i}+\sum_{i\in V}U_{i}n_{i}(n_{i}-1).
\end{equation}
Here $V$denotes the set of sites or vertices and edges of a Graph
$G=(V,E)$ and $E$ is the set of edges. $b_{i}^{\dagger}(b_{i})$
denote the spinless bosonic creation (annihilation) operators on site
$i$, $t_{ij}$ and $U_{i}$ are real parameters while $n_{i}=b_{i}^{\dagger}b_{i}$
is the particle number operator on site $i$.

Originally the model was independently proposed by Hubbard \cite{Hubbard1},
Kanamori \cite{Hubbard2} and Gutzwiler \cite{Hubbard3} for fermionic
systems and by Gersch and Knollman for bosonic systems \cite{Hubbard4}.
While it strongly simplifies the interactions in a real solid, by
reducing them to a hopping term with hopping strength $t_{ij}$ and
an on site interaction $U_{i}$, it already correctly predicts plenty
of effects in real solids including ferromagnetism \cite{LiebFerro,Mielke91a,saturated,decorated,MT93}
and superconductivity \cite{supercond} in the fermionic, and superfluid-insulator
transition \cite{superfluid} in the bosonic model. In many situations
one will choose a translation invariant graph, homogeneous on site
interaction and next-neighbour hopping (i.e. $U_{i}=U$ for all $i\in V$,
$t_{ij}=t$ for $\{i,j\}\in E$ and $\{i,j\}\in E$ iff $\vert i-j\vert=1$).
However, in the class of models under consideration in this paper
we will need two different hoppings and will not require any kind
of translational invariance as we will see later on. For more general
background information we refer to overviews by Lieb \cite{Overview1},
Tasaki \cite{Overview2} and Mielke \cite{Overview3}, since we will
focus our intention on the a certain subclass of Hubbard models: those
on line graphs with a flat band.

\subsection{Hubbard models on line graphs}

There are multiple ways of constructing Hubbard models with single
particle flat bands. One method is mentioned by Lieb \cite{LiebFerro}.
As he points out, for any bipartite graph $G$ consisting of the two
subgraphs $A$ and $B$ with $\vert B\vert=n\vert A\vert$ and $n\ge2$
the hopping matrix $T$ with $T_{ij}=t$ for $\{i,j\}\in E$ and 0
otherwise has at most rank $2\vert A\vert$ and therefore at least
$\vert V\vert-2\vert A\vert=\vert B\vert-\vert A\vert=(n-1)\vert A\vert$
zero eigenvalues. Consequently the system has $n-1$ flat bands in
the centre of the spectrum. He was able to give the first proof of
itinerant ferrimagnetism for these systems in case of a repulsive
interaction (i.e. $U_{i}>0$ for all $i$.) Although here ferromagnetism
only occurs in a weaker sense of the spin being an extensive quantity
and is in fact not saturated, this example already shows why Hubbard
models with flat bands are such interesting objects.

Of particular interest in the last years and focus of this work are
Hubbard models with flat bands at the bottom of the spectrum. One
class of graphs with such a low lying flat band is formed by decorated
lattices, as they are treated for example by Tasaki \cite{decorated},
a second class are line graphs of bipartite and two-connected graphs
\cite{saturated}.

As one can see from these examples, early research into graphs with
flat bands focused heavily on fermionic systems. However, recent development
has shown that bosons might be just as interesting to study on them,
as we will discuss later on. Since line graphs are at the heart of
the present work we will go over their construction. Our explanations
are based on those presented in \cite{MielkeHCB}, which we generalise
to also include toroidal graphs (i.e. graphs that can be drawn on
a torus without having any edges crossing another one) and to weaken
the condition on bipartition. We also refer to this work for greater
details on some aspects of the construction.

\begin{figure}[t]
\centering{}\includegraphics[width=12cm]{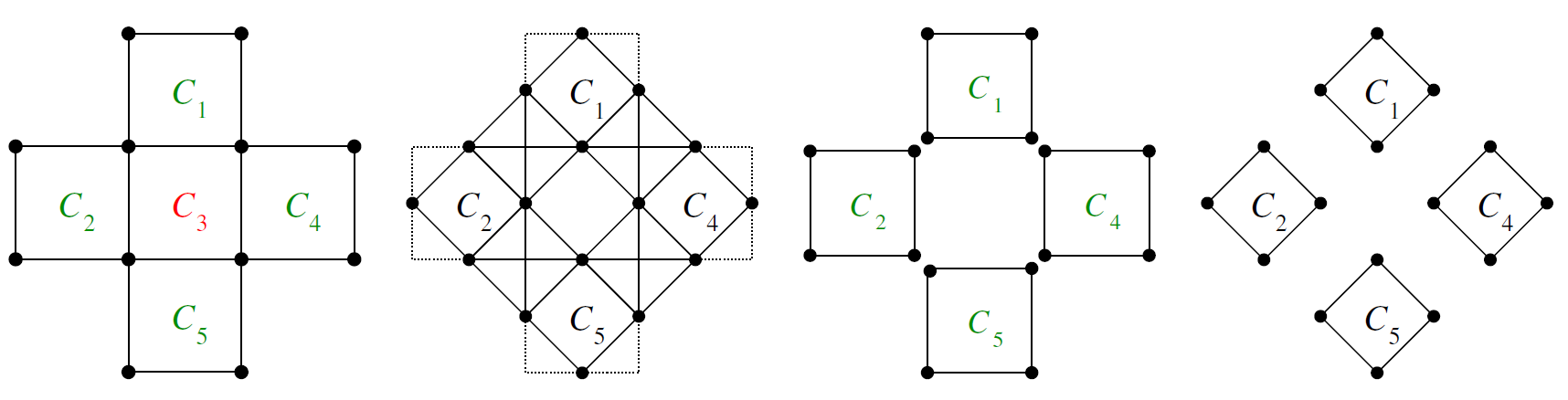} \caption{The left two images show a simple square lattice $G$ and its line
graph $L(G)$. The members of $F_{1}(G)$, $C_1$, $C_2$, $C_4$, and $C_5$ are coloured in green and
the right two images show the graph $G_{1}$ and its line graph $L(G_{1}).$
Figure taken from \cite{MielkeHCB}.}
\label{linegraph} 
\end{figure}

We start with some finite, toroidal and two-connected graph $G=(V(G),E(G))$.
It should be noted that the set of graphs that are both finite and
planar is a subset of toroid graphs, therefore this work applies to
them as well, while it also includes two dimensional graphs with doubly
periodic boundary conditions (DPBC), which are not planar. Once again
$V(G)$ and $E(G)$ denote the set of vertices and edges of $G$.
The line graph of $G$ is now given by $L(G)=(V(L(G)),E(L(G)))$ with
$V(L(G))=E(G)$ and $E(L(G))=\{\{e,e'\}\vert e,e'\in E(G),\vert e\cap e'\vert=1\}$.
For a more intuitive understanding, one can imagine the construction
of the line graph from the original graph in the following way: We
draw a vertex on each edge of the toroidal representation of $G$
and connect two vertices by an edge in $L(G)$ if and only if the
edges they are drawn upon have a common vertex in the original graph.
An illustration of the process for a simple square lattice can be
found in the left two images of figure (\ref{linegraph}). On this
line graph we can now define our Hamiltonian: 
\begin{equation}
H=\sum_{\{e,e'\}\in E(L(G))}\!\!t_{ee'}b_{e'}^{\dagger}b_{e}+\sum_{e\in V(L(G))}\!U_{e}n_{e}(n_{e}-1)
\end{equation}
Furthermore, we will only consider the hard-core limit $U_{e}\to\infty$
for all $e$ such that at maximum one particle can be placed on each
vertex in $L(G)$. Hence $H$ can be written as 
\begin{equation}
H=P_{\le1}\!\!\sum_{\{e,e'\}\in E(L(G))}\!\!t_{ee'}b_{e'}^{\dagger}b_{e}P_{\le1},
\end{equation}
where $P_{\le1}$ denotes the projector on the subspace of the Fock
space with at maximum one particle on each vertex.

Before we can define the hopping strengths on the line graph, we first
need to introduce some additional terms. The toroidal representation
of the original graph $G$ decomposes the plane into faces and since
a torus is a bounded object, all surfaces are bounded as well. We
call the set of all null-homotopic faces $F(G)$ and by $C\in F(G)$
we denote both the surface itself and its boundary. Since $G$ is
connected, there is at most one one-homotopic face. We call the elements
of $F(G)$ elementary cycles (they are indeed cycles since $G$ is
toroid.) Since the elementary cycles $C$ are subgraphs of $G$, we
can define their set of vertices $V(C)$ and their set of edges $E(C)$.
We now look at the colouring of the surfaces. We can colour two surfaces
$C$ and $C'$ with the same colour if they have no edge in common
(i.e. $E(C)\cap E(C')=\emptyset$).

Now let $F_{1}(G)\subset F(G)$ be the largest set of surfaces that
can be coloured with the same colour and that all have even cycle
length. Note that for a bipartite graph the second condition is trivially fulfilled for any subset of $F(G)$ since all cycles in bipartite
graphs have even length. If there are multiple of these sets, we just
choose one of them. By $E_{1}(G)$ we denote the set of edges, which
form the boundaries in $F_{1}(G)$, $E_{1}(G)=\bigcup_{C\in F_{1}(G)}E(C)\subset E(G)$.
Every edge in $E(G)$ belongs to a cycle in $F_{1}(G)$ if all faces
that are not in $F_{1}(G)$, including the potential one-homotopic
one, can be coloured by a single colour. Otherwise $E(G)\setminus E_{1}(G)$
is non-empty and we call its elements interstitials.

Additionally we define the graph $G_{1}$ which consists of all vertices
and edges from cycles in $F_{1}(G)$ where we consider the vertices
from different cycles to be distinct, even if they correspond to the
same vertex in G. (See figure (\ref{linegraph}) for an illustration
of the process.) Note that there is a one-to-one mapping between $E(G_{1})$
and $E_{1}(G)$ since every edge in $E_{1}(G)$ belongs to exactly
one cycle in $F_{1}(G)$ (otherwise they could not all be coloured
by the same colour) and therefore we write $E(G_{1})=E_{1}(G)$ while
acknowledging the slight imprecision of the expression.

The same can however not be said about $V(G_{1})$ and $V_{1}(G)=\bigcup_{C\in F_{1}(G)}V(C)$
since one vertex in $V_{1}(G)$ can be counted for multiple times
in $V(G_{1})$. Therefore, $V(L(G_{1}))=E(G_{1})=E_{1}(G)\subset E(G)=V(L(G))$
and two edges in $E(G_{1}),$ which are connected to a common vertex
are also connected in $E(G),$ hence $E(L(G_{1}))\subset E(L(G).$
Consequently we conclude that $L(G_{1})$ is a subgraph of $L(G)$
an we can finally write down our hopping strengths: 
\begin{align}
t_{ee'}=\begin{cases}
t & \textrm{if }\{e,e'\}\in E(L(G_{1}))\\
t' & \textrm{if }\{e,e'\}\in E(L(G))\setminus E(L(G_{1}))
\end{cases}
\end{align}
This means we allow for hopping strength $t$ between the edges of
elementary cycles and $t'$ on all other edges of the line graph.
Throughout this paper we choose $t>0$ and $t\ge t'\ge0$ and we will
note additional restrictions whenever they become necessary. Furthermore
it should be noted that since $G_{1}$ consists of isolated cycles,
it is isomorphic to its line graph, meaning that any cycle $C\in F_{1}(G)$
corresponds to a cycle of the same length in $L(G)$ and every edge
in $C\in F_{1}(G)$ corresponds to a vertex on the corresponding cycle
in the line graph. This allows to rewrite our Hamiltonian: 
\begin{equation}
H=tP_{\le1}\sum_{C\in F_{1}(G)}H_{C}P_{\le1}+t'P_{\le1}\sum_{\langle C,C'\rangle}H_{C,C'}P_{\le1}+t'H_{I}\label{Hamiltonian}
\end{equation}
Here $H_{C}$ describes the jumping on the elementary cycle $C$,
$H_{C,C'}$ the jumping between two neighbouring
cycles $C$ and $C'$ (i.e. $V(C)\cap V(C')\neq\emptyset$)
and $H_{I}$ the jumping to, from and between any interstitials. 

\subsection{Example: The chequerboard chain}

\begin{figure}
\centering{}\includegraphics[width=12cm]{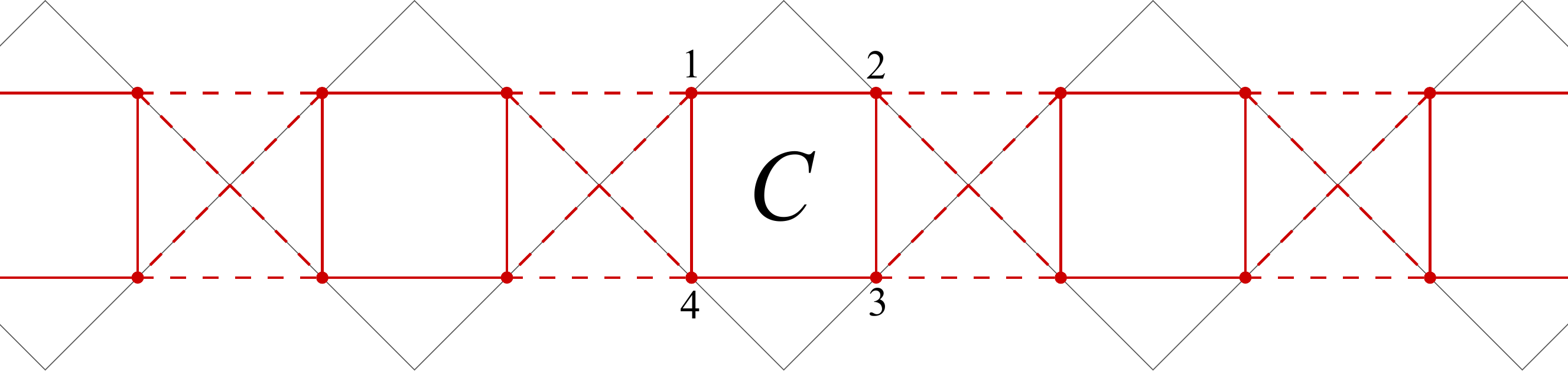} \caption{Illustration of the chequerboard chain: jumping on the cycles with
hopping strength $t$ is indicated by a continuous read line, jumping
between them with hopping strength $t'$ by a dashed red line and
for the graph of corner sharing graphs, whose line graph the chequerboardchain is, thin grey lines in the background are used.}
\label{chequerboard} 
\end{figure}

There are many different graphs within this class, but to get a more
concrete idea, we will have a closer look into one specific: the chequerboard
chain. It is defined as the line graph of a chain of corner sharing
squares and consists of a chain of cycles of length $\vert C\vert=4$,
which are connected by complete graphs. It is depicted in figure (\ref{chequerboard}).
As it was also explained in \cite{MielkeHCB}, we can construct compactly
localised one particle eigenstates $\psi_{C}$ with eigenvalue $-2t$
independently of $t'$ on any cycle $C$ in the line graph by labeling
an arbitrary vertex with the index $1$ and then we number them clockwise. Now we choose $\psi_{C}=\frac{1}{2}\!\left(\!b_{1}^{\dagger}(C)\!-\!b_{2}^{\dagger}(C)\!+\!b_{3}^{\dagger}(C)\!-\!b_{4}^{\dagger}(C)\!\right)\vert0\rangle$,
where $b_{i}^{\dagger}(C)$ is a bosonic creation operator on site
$i$ of cycle $C$; i.e. the absolute value is the same along the
cycle while the sign alternates. It is easy to see that they are indeed
eigenstates on $C$ with eigenvalue $-2t$ and the alternating signs
ensure that the jumping terms to neighbouring cycles vanish. For $t'<t$
these are unique ground states of the system. For $t'=t$ they remain
ground states, in case of periodic boundary conditions (PBC) they
are no longer unique though. As the eigenstates on different cycles
do not overlap, we can construct multi particle ground states from
them, simply by putting at maximum one particle in the ground state
of each cycle $C$. Again it turns out that these are the only ground
states for $t'<t$. This is possible until there are $N=\vert F_{1}(G)\vert$
particles in the system. We call this the critical density. 

\subsection{Ground states at or below the critical density\label{lowdensity}}

In fact this result holds for the general class of graphs we are looking
at. Mielke proved this for the class he was treating in \cite{MielkeHCB}
and the proof straightforwardly generalises to the larger class considered
here. The precise statements are:

\textbf{Proposition} \textit{Let $G=(V(G),E(G))$ be a finite two-connected
toroidal graph. Then for the Hubbard model on its line graph as defined
in (\ref{Hamiltonian}) the following holds: For $t>t'\ge0$ the ground
state eigenvalue of the single particle Hamiltonian is $-2t$ and
is $\vert F_{1}(G)\vert$-fold degenerate. For any $C\in F_{1}(G)$
a ground state is given by 
\begin{equation}
\psi_{C}=\frac{1}{\sqrt{\vert C\vert}}\sum_{i=1}^{\vert C\vert}(-1)^{i+1}b_{i}^{\dagger}(C)\vert0\rangle
\end{equation}
and its eigenvalue does not depend on $t'$.}

Here we have once again labeled an arbitrary edge of $C$ (corresponding
to a vertex in the line graph) with $1$ and then numbered the others
consecutively. Note that these are indeed eigenstates, since all cycles
in $F_{1}(G)$ have even length and any hopping to other edges cancels
out, since any edge in $G\setminus C$ that is connected to $C$ is
connected to exactly two neighbouring edges in $C$. The proof for
the larger class does not change compared to Mielke's proof: The $\psi_{C}$
are still minima of the first term in the rewritten Hamiltonian and
the only minima of the second term. Also it should be noted that these
states remain ground states in the homogeneous case ($t=t'$), although
they are no longer unique in general. More precisely, if $G$ is planar,
the ground states remain unique only if $F(G)\setminus F_{1}(G)$
contains no cycles of even length and if not, we need to additionally
demand that no cycle that is a boundary of the potential one-homotopic
face is of even length. Otherwise, additional linearly independent
ground states can be placed on the line graphs of the corresponding
cycles in the same manner as for the cycles in $F_{1}(G).$ This also
explains, why for the homogeneous chequerboard chain the ground states
remain unique for open boundary conditions, as its underlying graph
is planar and $F(G)\setminus F_{1}(G)=\emptyset.$ On the other hand,
for PBC the square chain surrounds either another face in $F(G)$
of even length if a planar representation is chosen or, if a non planar
representation is chosen, there are two cycles of even length at the
boundary of the one-homotopic face (for either of which the ground
state on them can be added to the ones of the inhomogeneous case,
to create a base of the ground states of the homogeneous system).
For the multi particle ground states at or below the critical density
the result is given by:

\textbf{Proposition} \textit{For $t>t'>0$ and $N\le\vert F_{1}(G)\vert$,
the ground states of the Hubbard model on line graphs of finite two-connected
toroidal graphs with $N$ hard core bosons are the same as the ones
for $t'=0$ and the ground state energy is $-2tN.$}

The proof also generalises directly from Mielke's proof to the broader
class in question, since the eigenstates of the $t'=0$ case remain
eigenstates of the full Hamiltonian and are minima of the first term
and the only minima of the second. Once again in the homogeneous case
these states remain ground states but are no longer unique in general.
For more information on the homogeneous case at or below the critical
density, we refer to \cite{homogeneous}.

\section{Hard core bosons above the critical density}

As we have seen, we are able to characterize the ground states up
to the critical density. The naturally occurring question is what
happens if we add an additional particle beyond the critical density.
To at least give a partial answer to that question, we need to establish
some additional constraints on the class of graphs we are treating.
First of all, we demand that there are no interstitials present on
the line graph $L(G)$. This is equivalent to $\bigcup_{C\in F_{1}(G)}E(C)=E(G)$
and, as we have seen, can also be expressed as it being possible to
colour the original graph $G$ (including the potential the potential
one-homotopic face) where $F_{1}(G)$ contains all members of one
colour. Additionally, we demand all cycles $C$ in $F_{1}(G)$ to
be of length $4$. It is important to note that all other faces in
$F(G)$ might have boundaries of arbitrary length.

Intuitively one can imagine the construction of an arbitrary graph
in this class as follows: We start with a simple torus, which we colour
in one colour, e.g. yellow, and then place a single quadrilateral on
it, which we w.l.o.g. colour in blue. Now we continue to place additionalblue quadrilaterals on the torus which need to obey the following
conditions: they need to be connected to the other quadrilaterals
(i.e. if we number the quadrilaterals according to the order in which
we place them on the torus and denote them by $C_{k}$, then for any
$k$ there needs to be some $1\le i<k$ such that $V(C_{k})\cap V(C_{i})\neq\emptyset$).
It must not share an edge with any of the quadrilaterals (i.e. $E(C_{k})\cap E(C_{i})=\emptyset$
for $i\neq k$; this allows for quadrilaterals to have the same colour).
Last they must not intersect with any of the previously placed quadrilaterals
(i.e. leave the toroidal structure of the graph intact.) Now one repeats
this process arbitrarily often and ends up with a graph that is toroidal, has only two colours and where the blue surfaces make up the elements
of $F_{1}(G)$ and the potential remaining one-homotopic surface is
coloured in yellow. Therefore, this graph has no interstitials, is two-connected
(since removing any edge from a quadrilateral still allows for another
path along the other three edges of the quadrilateral) and toroidal
by construction.

\begin{SCfigure}
\caption{Example of a non-bipartite graph $G$ within the class of graphs, to which the main result applies. It includes cycles of length 3 (the
yellow triangles), but since they do not appear among the surfaces of
$F_{1}(G)$ (coloured in blue), we can still construct localised
eigenstates as described in section \ref{lowdensity} on the graph.}
\includegraphics[width=7.2cm]{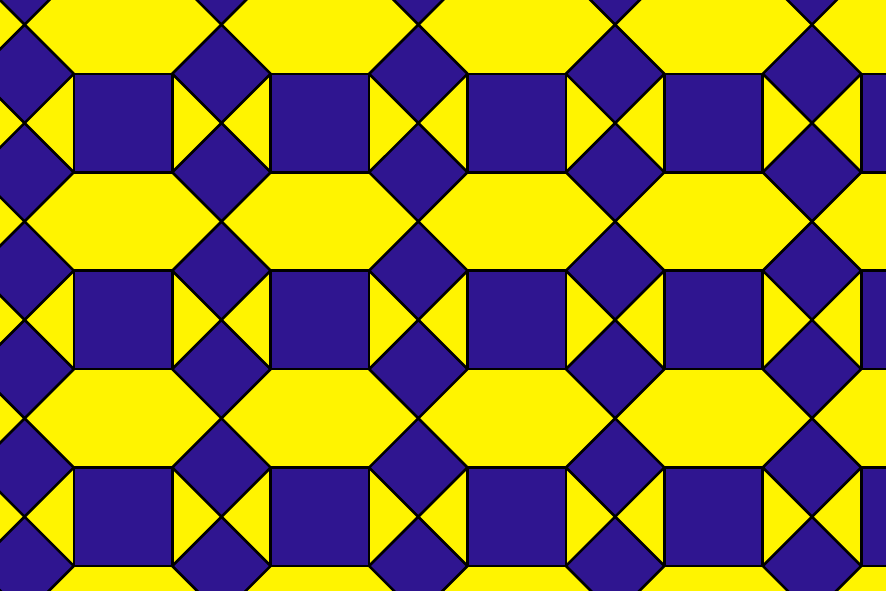} \label{nonbipartite}
\end{SCfigure}

Both the one dimensional chequerboard chain and its two dimensional
analogue, the chequerboard lattice, are a member of this class for
arbitrary boundary conditions. So are all other examples mentioned
in figure \ref{localsymmetry} (for the regularly shaped ones once
again regardless of boundary conditions) and figure \ref{nonbipartite}
features a member of the class, which is not bipartite. Other interesting cases like the kagome lattice are however not, as its cycles have
length six and three colours are necessary to colour its underlying
graph, the honeycomb lattice. The limitation to cycles of length $4$
could be avoided without much theoretical difficulty. However, the
results become generally worse with longer cycles, as it is energetically less expensive to put multiple particles into one cycle and mixed
cycle lengths would require additional treatment. The reason we are
excluding the existence of interstitials is that they allow for additional
possibilities for the additional particle to spread to.

Under these conditions our initial question (What would happen if
there is an additional particle placed on the line graph?) has an
obvious answer for $t'=0$: The additional particle can simply be
put on any cycle $C\in F_{1}(G)$ to form the two particle ground
state (which has energy $-2\sqrt{2}t$, see table \ref{tableHCB}
for its precise form) and still have one particle each in the single
particle ground state on each other cycle, we call these ground states
$\varphi_{C}$ and they span any ground state for $t'=0.$ We denote
this subspace by $\Omega_{0}$ and the projector onto it by $P_{0}$.
Since the two particle one cycle ground state is no longer an eigenstate
for $t'\neq0,$ the result for the uncoupled system can no longer
be as easily generalised to the full Hamiltonian. We were however
able to derive rigorous boundaries for the energy of the lowest states,
to show how they remain separated from the rest of the spectrum and
how they are dominated by the ground states of the uncoupled system.

Before we get to it, we however need to quote one more result on the
chequerboard chain by Drescher and Mielke, which we will also extend
to some additional graphs, to understand the full importance of our
result for these specific graphs.

\subsection{Degeneracy of ground states of graphs possessing a local reflection
symmetry}

We look at the chequerboard chain with PBC. Using a symmetry argument,
Drescher and Mielke were able to show the following result, which
we slightly rephrase to cater towards the needs of this work \cite{degeneracy}:

\textbf{Proposition (Drescher, Mielke, 2017)} \textit{If for some
ground state $\varphi_{1}$ of $H$ on the chequerboard chain with
PBC, there is an overlap with the ground states of the uncoupled system
(i.e. $\Vert P_{0}\varphi\Vert_{2}\neq0$), then the ground states
are at least $\vert F_{1}(G)\vert$ fold degenerated.}

This can be proved using a symmetry property of the chequerboard chain:
Its Hamiltonian remains invariant under exchanges of the lower vertices
with their upper counterparts (i.e. in figure \ref{chequerboard}
to interchange vertex 1 with 4 and 2 with 3, respectively, or, equivalently,
to exchange the upper and lower edges of the square in the original
graph) for any cycle $C$ and we call the operator that performs this
operation $S_{C}$. Therefore, it is possible to find a common base
of eigenstates of $H$ and all of the $S_{C}$ (as they also commute
with one another.) Since the base states of $\Omega_{0},$ $\varphi_{C'}$
have eigenvalue $2\delta_{C,C'}-1$ under $S_{C},$ there is a unique
signature for all base states. Now we look at a ground state $\varphi_{1}$
of H with $\Vert P_{0}\varphi_{1}\Vert_{2}\neq0$ which is also an
eigenstates of all $S_{C}.$ Therefore, its signature needs to be
identical to one of the base states of $\Omega_{0}$. Additionally,
the Hamiltonian is invariant under shifting the state by one face
to the right due to the PBC. Hence, the resulting state $\varphi_{2}$
must also be a ground state but since its signature differs from the
one of $\varphi_{1}$ they cannot be identical. This argument can
be repeated to create $\vert F_{1}(G)\vert$ ground states $\varphi_{k}$
with unique signatures, which therefore are linearly independent,
which proves the proposition. Moreover the $\varphi_{k}$ only overlap
with a single base state of $\Omega_{0}$, the one they share the
signature with. Hence they obey $P_{0}\varphi_{k}=r\varphi_{C}$,
for some $C\in F_{1}(G)$ and $0<\vert r\vert\le1$, i.e. their projector
into into $\Omega_{0}$ has a compactly localised pair.

\begin{figure}
\centering \begin{subfigure}[b]{0.3\textwidth} \includegraphics[width=1\textwidth]{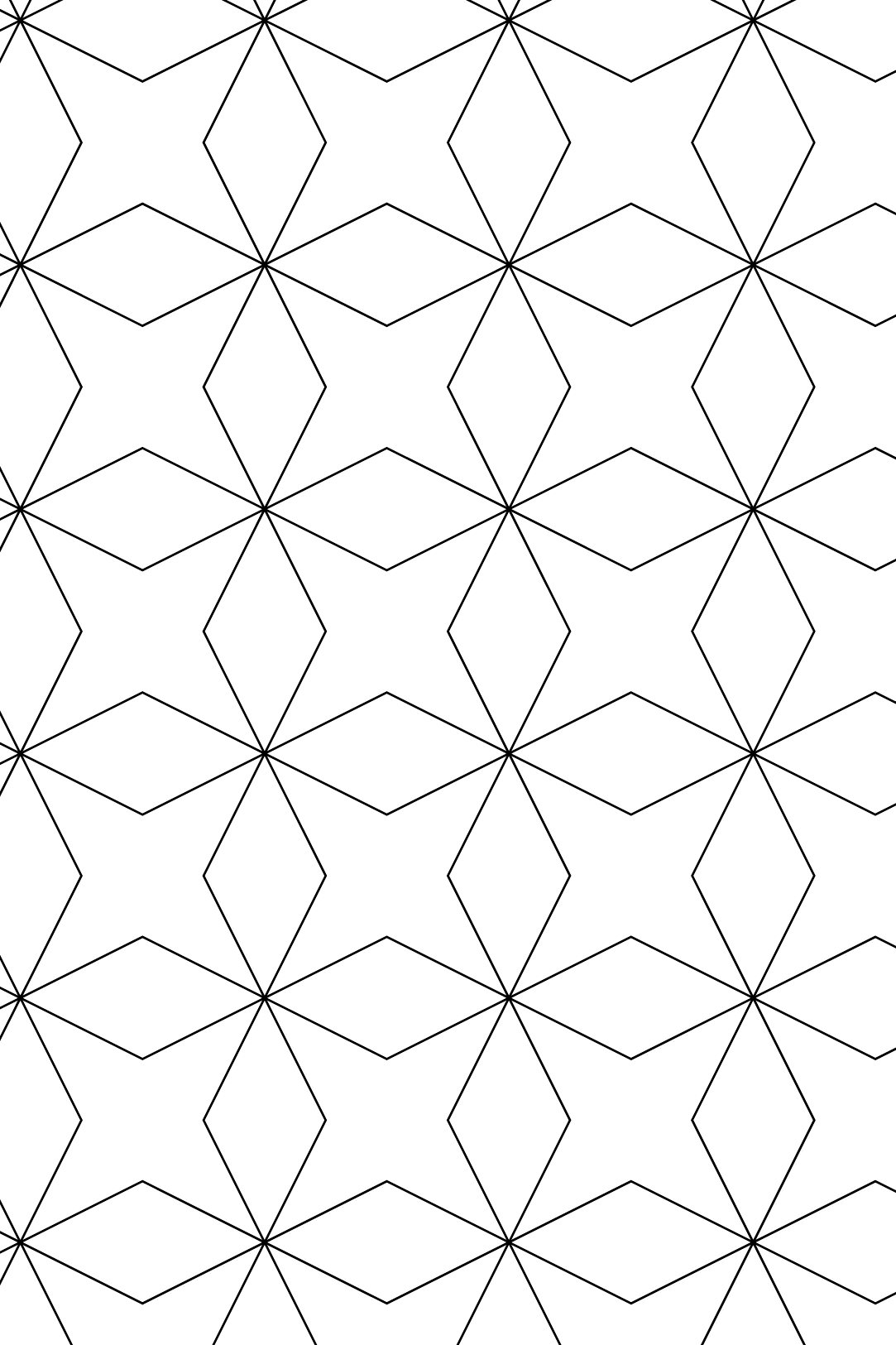}
\caption{The LIR (4,1)}
\label{LIR41} \end{subfigure} \begin{subfigure}[b]{0.3\textwidth}
\includegraphics[width=1\textwidth]{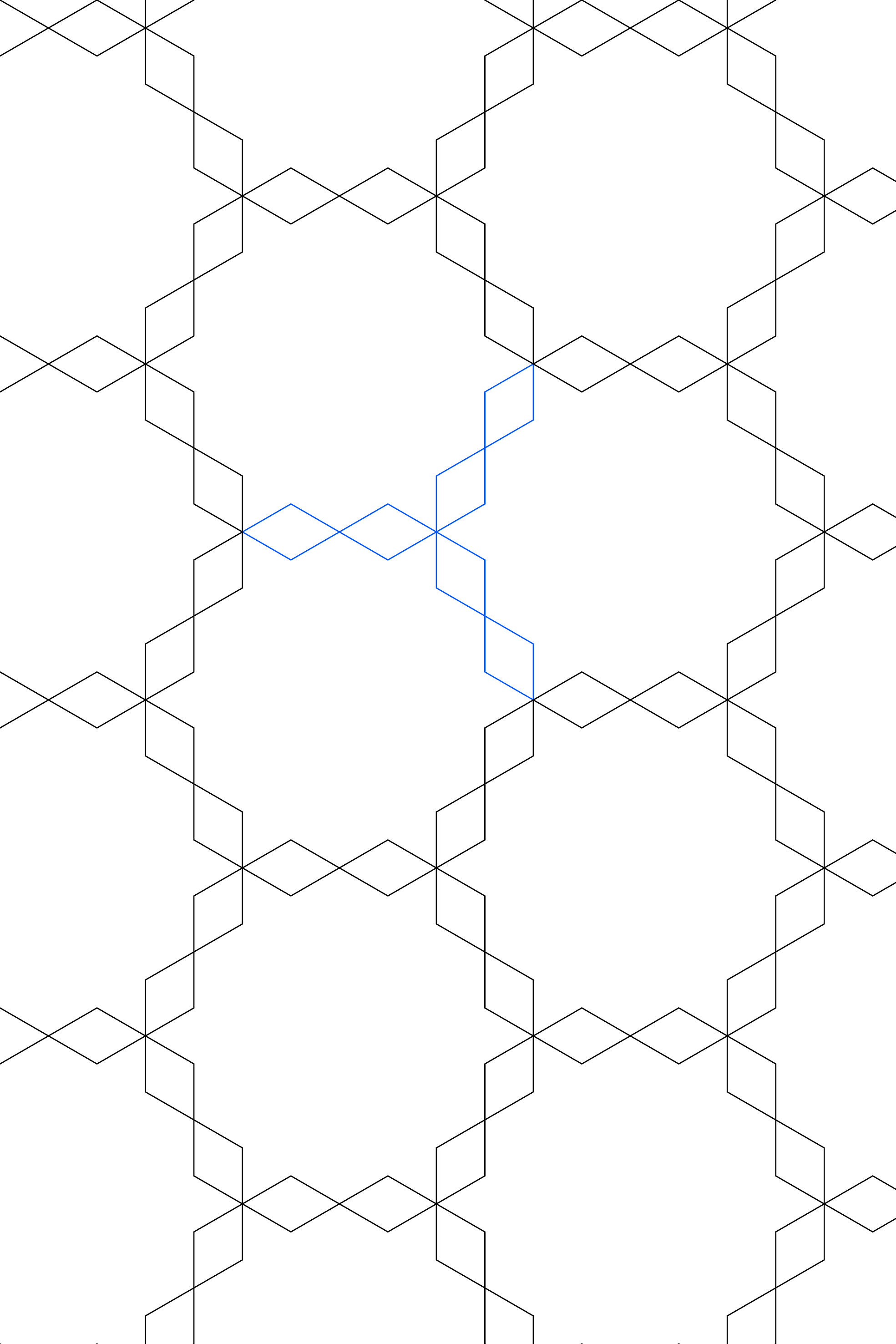} \caption{The LIR(3,2)}
\label{LIR32} \end{subfigure} \begin{subfigure}[b]{0.3\textwidth}
\includegraphics[width=1\textwidth]{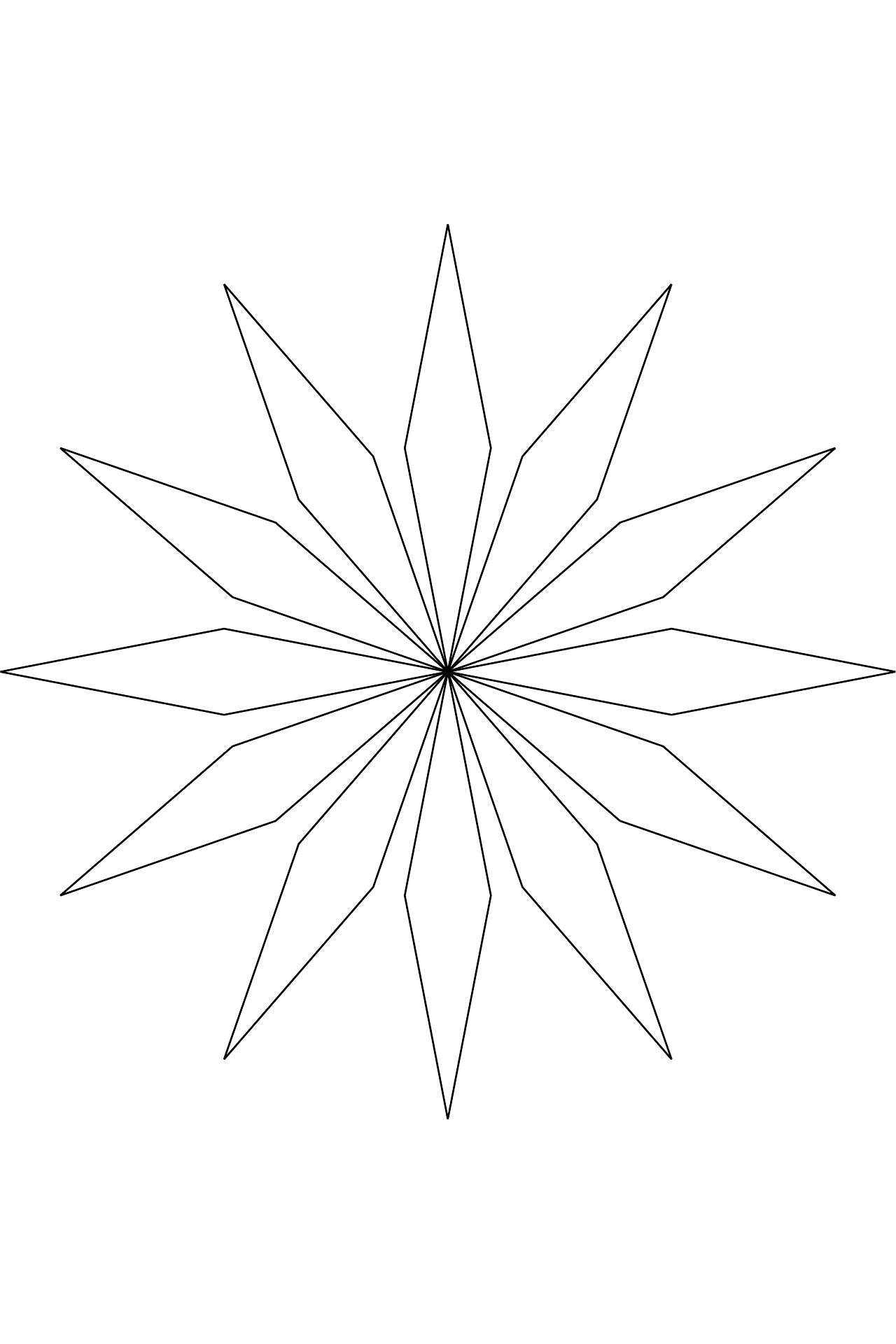}
\caption{The 12 rhombi star }
\label{rhombistar} \end{subfigure} \caption{Several toroidal graphs obeying a local reflection symmetry.}
\label{localsymmetry} 
\end{figure}

We now look at the lattice of intertwined rhombi chains (abbreviated
LIR (4,1), here 4 stands for the maximum number of rhombi that share
a single vertex and 1 for the number rhombi between two such vertices)
with DPBC as depicted in figure \ref{LIR41}). Since the rhombi share
only corners but not edges, they can indeed be coloured by one colour
and they form the set $F_{1}(G).$ One immediately notices that its
line graph does possess the same local reflection symmetry, since
for all rhombi the mirroring on the longer diagonal leaves the Hamiltonian
of the line graph invariant and we can again find a common base of
eigenstates of $H$ and all $S_{C}.$ Any ground state of $H$ in
this basis can overlap with at maximum one base state of $\Omega_{0}.$
Now we assume such an overlapping ground state, $\varphi_{1},$ does
indeed exist and $\varphi_{C}$ is the state it overlaps with. Then
we can again use the translational invariance due to the DPBC to find
additional ground states which overlap with different base states
of $\Omega_{0}.$ This is possible for all $\varphi_{C}'$ as long
as $C$ and $C'$ share the same spatial orientation (i.e. are either
both members of a horizontal or both members of a vertical chain.)
Now we additionally demand that the lattice has the same length in
both directions and call this the symmetric case. The Hamiltonian
is then also invariant under a global rotation by $\frac{\pi}{2}$
around an arbitrary vertex of $G$ which connects multiple rhombi.
This maps $C$ onto a rhombus of different orientation and therefore
transforms $\varphi_{k}$ onto a ground state overlapping with a state
in $\Omega_{0}$ with two particles on that rhombus. Repeating the
translations we can find a ground state of $H$ which exactly overlaps
with $\varphi_{\bar{C}}$ for every $\bar{C}\in F_{1}(G).$

Similarly one can work out the ground state degeneracy under this
condition for the line graph of the thinned out lattice of intertwined
rhombi (abbreviated LIR (3,2), use of the numbers in accordance to
the (4,1) case.) It is depicted in figure \ref{LIR32}). Starting
from one ground state with a given signature we can use the translational invariance to find linearly independent ground states with signature
$+1$ on every rhombus in the lattice, given it has the same position
in the "rotor" the graph is build from (marked in blue in fig \ref{LIR32}).)
The rhombi enclose regular hexagons which can be thought of as a honeycomb
lattice, if we consider the rhombi to be their edges. If all chains
of hexagons have the same length, we call this again the symmetric
case. In this case, the Hamiltonian is invariant under rotations by
$\frac{\pi}{3}$ around the centre of any "hexagon" which allows
us to find find ground states with signature $+1$ on every rhombus
of the "basic rotor" and in combination with the translational invariance,
we achieve the same result as for the chequerboard chain:

\textbf{Proposition} \textit{If the line graphs of the symmetric LIR
(4,1) with DPBC or of the symmetric LIR (3,2) with DPBC possess a
ground state $\varphi_{1}$ of $H$ that has an overlap with the ground
states of the uncoupled system (i.e. $\Vert P_{0}\varphi\Vert_{2}\neq0$),the ground states are at least $\vert F_{1}(G)\vert$ fold degenerated.}

\textit{Remarks:} Several similar regular two dimensional lattices
obeying this local symmetry can be found. We decided to present these
two specifically because the (4,1) version seems like the most intuitive
one, while in the (3,2) case every rhombus only neighbours three others,
which is the minimal possible number for a regular two dimensional
structure which proves to be beneficial for the main result.

Given a sufficiently small acute angle, one can of course attach arbitrarily
many rhombi to a central vertex and we call this the $n$ rhombi star.
Essentially the LIR are build up from such stars for $n=4$ and $n=3$, respectively. For $n=12$ it is depicted in figure \ref{rhombistar}).
Since it obeys the local reflection symmetry and is invariant under
global rotations by $\frac{2\pi}{n}$ around its centre, an analogue
proof for the degeneracy of its ground state energy is possible. However,since the number of other next neighbours is proportional to the number
of rhombi within the graph, the main result very quickly gets extremely
bad even for a moderate number of rhombi (and therefore allowed particles in the systems).
Hence, we consider the (isolated) star with an arbitrary $n$ to be more of an academic curiosity, than a physically relevant
system (at least in the context of localised pair formation).

While the energetic degeneracy depends on some form of a global symmetry,the possibility to find ground states that overlap with at maximum
one base state of $\Omega_{0}$ only relies on the existence of the
local reflection symmetry for all $C.$ Therefore, it also holds for
all other graphs in our class obeying this symmetry. Examples of graphs
obeying the local symmetry but without any kind of global rotational
or translational invariance are treelike structures such as the one
described in chapter 5 of \cite{MielkeHCB}.

\subsection{Main result}

Under the conditions set out at the beginning of this chapter we now
show the following rigorous bounds for the energy of the lowest energy
states of $H$ as well as for the gap to the next lowest states up
to a certain $t'_{\sup}$ and show that they are dominated by states
in $\Omega_{0}.$

\textbf{Theorem} \textit{Let $G$ be a two-connected toroidal graph
with $\bigcup_{C\in F_{1}(G)}E(C)=E(G)$ and $\vert C\vert=4$ for
all $C\in F_{1}(G)$. Then the following holds for $t'<t'_{\sup}$:}
\begin{enumerate}
\item \textit{The Hamiltonian 
\begin{equation}
H=tP_{\le1}\sum_{C\in F_{1}(G)}H_{C}P_{\le1}+t'P_{\le1}\sum_{C\neq C'\in F_{1}(G)}H_{C,C'}P_{\le1}
\end{equation}
with $N=\vert F_{1}(G)\vert+1$ hard core bosons on $L(G)$ has exactly
$\vert F_{1}(G)\vert$ linearly independent eigenstates with energies
$E_{k}$ in the lowest interval, sufficing 
\begin{equation}
-\left(2N-e_{5}\right)t-b\sqrt{c(G)}\eta_{+}(t')t'\le E_{k}\le-\left(2N-e_{5}\right)t.
\end{equation}
}
\item \textit{The energy gap $\Delta$ between these lowest and the upper
states satisfies 
\begin{equation}
\Delta\geq e_{5}t-\left(a\frac{t'}{t}c^{2}(G)+b\eta_{-}(t')\sqrt{c(G)}\right)t'.
\end{equation}
.}
\item \textit{The corresponding eigenstates $\varphi_{k}$ satisfy 
\begin{align}
\Vert P_{0}\varphi_{k}\Vert_{2}\ge\left(1+\eta_{-}^{2}(t')\right)^{-\frac{1}{2}}\Vert\varphi_{k}\Vert_{2}.
\end{align}
} 
\end{enumerate}
\textit{Where 
\begin{align}
\begin{split}t'_{\sup} & =\left(\sqrt{\frac{b^{2}}{a^{2}c(G)}+\frac{e_{5}}{a}}-\frac{b}{a\sqrt{c(G)}}\right)\frac{t}{c(G)}\\
 & =0.6120\left(\sqrt{\frac{0.08478}{c(G)}+1}-\frac{0.2912}{\sqrt{c(G)}}\right)\frac{t}{c(G)},
\end{split}
\\
\begin{split}\eta_{\pm}(t')= & \dfrac{2b\sqrt{c(G)}}{\left(e_{5}-a\tilde{t}^{2}c^{2}(G)\right)+\sqrt{\left(e_{5}-a\tilde{t}^{2}c^{2}(G)\right)^{2}\pm4b^{2}\tilde{t}^{2}c(G)}}\tilde{t}\\
= & \dfrac{b\sqrt{c(G)}}{e_{5}}\tilde{t}+\mathcal{O}\left(\tilde{t}^{3}\right),
\end{split}
\end{align}
$\tilde{t}=\frac{t'}{t}.$ $e_{5}$, $a$ and $b$ are numerical constants:\begin{align*}
e_{5} & =2\left(2-\sqrt{2}\right)=1.1714\\
a & =\frac{1425+974\sqrt{2})}{896}=3.1277\\
b & =\sqrt{\frac{3}{4}\left(\sqrt{2}-1\right)}=0.5574
\end{align*}
$P_{0}$ denotes the projector into the $N$ particle ground states
of 
\begin{equation}
H_{F_{1}(G)}=tP_{\le1}\sum_{C\in F_{1}(G)}H_{C}P_{\le1}
\end{equation}
and $\Vert\psi\Vert_{2}=\left(\sum_{i=1}^{n}\vert\psi_{i}\vert^{2}\right)^{\frac{1}{2}}$
the standard euclidean norm on the $N$ particle Fock space w.r.t.
the base of eigenstates of $H_{F_{1}(G)}$.}

\textit{Proof:} For our proof we will make use of a generalised version
of of Gershgorin's circle theorem by Feingold and Varga \cite{FeingoldVarga}.
They showed the following: Let $A$ be a square matrix which is acting
on a space $\Omega$ and is partitioned in the following manner: 
\begin{equation}
A=\begin{pmatrix}A_{11} & A_{12} & \dots & A_{1m}\\
A_{21} & A_{22} & \dots & A_{2m}\\
\vdots & \vdots & \ddots & \vdots\\
A_{m1} & A_{m2} & \dots & A_{mm}
\end{pmatrix}
\end{equation}
Where the $A_{ii}$ are square matrices acting on $n_{i}$ dimensional
subspaces $\Omega_{i}$ of $\Omega$, therefore $\sum_{i=1}^{m}n_{i}=n=:\dim(\Omega)$
and the $A_{ij}$ are $n_{i}\times n_{j}$ matrices. Then for every
eigenvalue $\lambda$ of $A$ there is at least one $i$, $1\le i\le m$
such that: 
\begin{equation}
\left(\left\Vert \left(A_{ii}-\lambda I_{i}\right)^{-1}\right\Vert \right)^{-1}\le\sum_{k=1,k\neq i}^{m}\left\Vert A_{ik}\right\Vert 
\end{equation}
Here $I_{i}$ denotes the unit matrix on $\Omega_{i}$ and the matrix
norms are the ones derived by the corresponding vector norms: 
\begin{equation}
\left\Vert A_{ij}\right\Vert =\max_{\substack{x\in\Omega_{j}\\
\Vert x\Vert_{\Omega_{j}}=1
}
}\left\Vert A_{ij}x\right\Vert _{\Omega_{i}}
\end{equation}
It is important to note that the vector norms on the $\Omega_{i}$
can be chosen arbitrarily and independent of each other. Also, since
\begin{equation}
\left(\left\Vert B^{-1}\right\Vert \right)^{-1}=\min_{\Vert x\Vert=1}\left\Vert Bx\right\Vert 
\end{equation}
for any invertible matrix $B$, it is the natural continuation to
define $\left(\left\Vert B^{-1}\right\Vert \right)^{-1}=0$ for singular$B$.

Once again let $\Omega_{0}$ be the space spanned by the ground states
$\varphi_{C}$ of $H_{F_{1}(G)}$, $\Omega_{1}$ its orthogonal subspace
and $P_{1}=1-P_{0}$ the corresponding projector. Then we can write
H as 
\begin{equation}
H=\begin{pmatrix}H_{0} & H_{01}\\
H_{10} & H_{1}
\end{pmatrix}=\begin{pmatrix}P_{0}HP_{0} & P_{0}HP_{1}\\
P_{1}HP_{0} & P_{1}HP_{1}
\end{pmatrix}.
\end{equation}
Furthermore we choose our norms as $\Vert\cdot\Vert_{\Omega_{0}}=\eta\Vert\cdot\Vert_{2}$
and $\Vert\cdot\Vert_{\Omega_{1}}=\Vert\cdot\Vert_{2}$ where $\Vert\cdot\Vert_{2}$
is the euclidean norm on $\Omega_{i}$, $i=0,1$ and $\eta$ a positive
real number, which we will specify later on.

Now our goal is to show that 
\begin{align}
\begin{split}\sup\Lambda_{0}:= & \sup\Big\{\lambda\in\mathbb{R}\Big\vert\left(\left\Vert \left(H_{0}-\lambda I_{0}\right)^{-1}\right\Vert \right)^{-1}\le\left\Vert H_{01}\right\Vert \Big\}\\
< & \inf\Big\{\lambda\in\mathbb{R}\Big\vert\left(\left\Vert \left(H_{1}-\lambda I_{1}\right)^{-1}\right\Vert \right)^{-1}\le\left\Vert H_{10}\right\Vert \Big\}=:\inf\Lambda_{1}\label{separation}
\end{split}
\end{align}
for all $t'<t'_{\sup}$ and an appropriate choice of $\eta$. Since
we know that for $t'=0$ there are exactly $\vert F_{1}(G)\vert$
eigenvalues of $H$ in $\Lambda_{0}$ and the eigenvalues vary continuouslyin $t'$, this number cannot change as long as the two sets are disjoint.
Therefore, we will have proven the separation of the lowest eigenvalues
from the others and will also be able to establish an upper boundary
for their values.

We start by looking at $\Lambda_{0}$. Since $H_{0}=\left(-2(N-2)-2\sqrt{2}\right)tI_{0}$
we have 
\begin{equation}
\left(\left\Vert \left(H_{0}-\lambda I_{0}\right)^{-1}\right\Vert \right)^{-1}=\left\vert \left(-2(N-2)-2\sqrt{2}\right)t-\lambda\right\vert .\label{H0}
\end{equation}
Also 
\begin{align}
\begin{split}\Vert H_{01}\Vert & =\max_{\substack{\psi\in\Omega_{1}\\
\Vert\psi\Vert_{\Omega_{1}}=1
}
}\Vert H_{01}\psi\Vert_{\Omega_{0}}\\
 & =\eta\sqrt{\rho\left(H_{01}^{\dagger}H_{01}\right)}\\
 & =\eta\sqrt{\rho\left(H_{01}H_{01}^{\dagger}\right)}\\
 & =\eta\sqrt{\rho\left(H_{01}H_{10}\right)},
\end{split}
\label{H01}
\end{align}
where $\rho(\cdot)$ is the spectral radius of a matrix. The third
equality holds because the spectral radius is invariant under Hermitian
transposition and the last one due to $H$ being Hermitian. One finds
the eigenvectors of $H_{01}H_{10}$ to be the same as the eigenvectors
of $H_{0}$ and their eigenvalues to be $\lambda_{k}=(\frac{3}{4}(\sqrt{2}-1))c(C_{k})t'^{2}=:b^{2}c(C_{k})t'^{2}$.
Where $C_{k}$ is the cycle occupied by two particles in the $k$th
eigenstate of $H_{0}$. Hence we can conclude 
\begin{equation}
\Vert H_{01}\Vert=\eta b\sqrt{c(G)}t'.
\end{equation}
Combining (\ref{H0}) and (\ref{H01}) we get 
\begin{equation}
\Lambda_{0}=\Big\{\lambda\in\mathbb{R}\Big\vert\;\left\vert \left(-2(N-2)-2\sqrt{2}\right)t-\lambda\right\vert \le\eta b\sqrt{c(G)}t'\Big\}
\end{equation}
and subsequently 
\begin{equation}
\sup\Lambda_{0}=\left(-2(N-2)-2\sqrt{2}\right)t+\eta b\sqrt{c(G)}t'.\label{sup}
\end{equation}
For $\Lambda_{1}$ the calculation is not as straight forward. $H_{1}$
is non diagonal and therefore we need to look at $\big(\big\Vert\left(H_{1}-\lambda I_{1}\right)^{-1}\big\Vert\big)^{-1}$
more carefully. Let $S$ be a diagonal Matrix with $S_{kk}>0$ for
all $k$. Then $S$ is non-singular and we can apply proposition 1.1
of \cite{matrixnorm} to achieve the following lower bound 
\begin{align}
\begin{split} & \left(\left\Vert \left(H_{1}-\lambda I_{1}\right)^{-1}\right\Vert \right)^{-1}\\
\geq & \left(\left\Vert \left(S\left(H_{1}-\lambda I_{1}\right)S^{-1}\right)^{-1}\right\Vert _{1}\left\Vert \left(S^{-1}\left(H_{1}-\lambda I_{1}\right)S\right)^{-1}\right\Vert _{\infty}\right)^{-\frac{1}{2}}\\
= & \left(\left\Vert \left(S\left(H_{1}-\lambda I_{1}\right)S^{-1}\right)^{-1}\right\Vert _{1}\right)^{-1}\\
= & \min_{\Vert\psi\Vert_{1}=1}\left\Vert S\left(H_{1}-\lambda I_{1}\right)S^{-1}\psi\right\Vert _{1}\\
= & \min_{\Vert\psi\Vert_{1}=1}\sum_{i=1}^{n_{1}}\Big\vert\left(S(H_{1}-\lambda I_{1})S^{-1}\right)_{ii}\psi_{i}+\sum_{j\neq i}\left(SH_{1}S^{-1}\right)_{ij}\psi_{j}\Big\vert\\
= & \min_{\Vert\psi\Vert_{1}=1}\sum_{i=1}^{n_{1}}\Big\vert(H_{1;ii}-\lambda)\psi_{i}+\sum_{j\neq i}S_{ii}H_{1;ij}S_{jj}^{-1}\psi_{j}\Big\vert\\
\ge & \min_{\Vert\psi\Vert_{1}=1}\sum_{i=1}^{n_{1}}\Big(\vert H_{1;ii}-\lambda\vert\vert\psi_{i}\vert-\sum_{j\neq i}\vert S_{ii}H_{1;ij}S_{jj}^{-1}\vert\vert\psi_{j}\vert\Big)\\
= & \min_{\Vert\psi\Vert_{1}=1}\sum_{i,j=1}^{n_{1}}\vert S_{ii}H_{1;ij}S_{jj}^{-1}-\lambda\delta_{ij}\vert\vert\psi_{j}\vert(2\delta_{ij}-1)\\
= & \min_{\Vert\psi\Vert_{1}=1}\sum_{j=1}^{n_{1}}\vert\psi_{j}\vert\Big(\vert H_{1;jj}-\lambda\vert-\sum_{i\neq j}\vert S_{ii}H_{1;ij}S_{jj}^{-1}\vert\Big)\\
= & \min_{1\le k\le n_{1}}\Big(\vert H_{1;kk}-\lambda\vert-\sum_{i\neq k}\vert S_{ii}H_{1;ik}S_{kk}^{-1}\vert\Big)\min_{\Vert\psi\Vert_{1}=1}\sum_{j=1}^{n_{1}}\vert\psi_{j}\vert\label{analogy}
\end{split}
\\
= & \min_{1\le k\le n_{1}}\Big(\vert H_{1;kk}-\lambda\vert-\sum_{i\neq k}\vert S_{ii}H_{1;ik}S_{kk}^{-1}\vert\Big).\label{LB}
\end{align}
We are interested in $\inf\Lambda_{1}$ and $\lambda=\left(-2(N-4)-4\sqrt{2}\right)t$
fulfils 
\begin{equation}
\min_{1\le k\le n_{1}}\Big(\vert H_{1;kk}-\lambda\vert-\sum_{i\neq k}\vert S_{ii}H_{1;ik}S_{kk}^{-1}\vert\Big)\le\left\Vert H_{10}\right\Vert 
\end{equation}
trivially for all $t',$ as it is the lowest non ground state eigenvalue
of $H_{F_{1}(G)}$. Therefore, we now assume $\lambda\le\left(-2(N-4)-4\sqrt{2}\right)t$
and note $\lambda\le H_{1;kk}$ for all $1\le k\le n_{1}$. Consequently
we can drop the absolute value in the first term without changing
its value. Our next goal is to rewrite the argument of (\ref{LB})
as a constant independent of $k$ plus a sum over the connections
between neighbouring cycles. To achieve this we introduce $e_{i}$,
the difference between the energy $\frac{E_{i}}{t}$ of the $i$th
one cycle eigenstate (normalised by $t$) and the again normalised
energy of distributing all $N_{i}$ particles in the state in one
particle ground states; 
\begin{equation}
e_{i}=\frac{E_{i}}{t}+2N_{i}.
\end{equation}
If the cycle $C$ is occupied by the $i$th one cycle eigenstate in
$\chi_{k}$, the $k$th base state of $H_{1},$ we then write $N_{k}(C)=N_{i}$,
$E_{k}(C)=E_{i}$ and $e_{k}(C)=e_{i}$, which allows us to rewrite
$H_{1;kk}$ as 
\begin{align}
\begin{split}H_{1;kk} & =\sum_{C}E_{k}(C)\\
 & =\sum_{C}\left(e_{k}(C)-2N_{k}(C)\right)t\\
 & =-2Nt+\sum_{C}e_{k}(c)t,\label{Diag}
\end{split}
\end{align}
since $\sum_{C}N_{k}(C)=N$ for all $k$ by definition. The values
of all $e_{i}$ are given in table \ref{tableHCB}. 
\begin{table}
\global\long\def\arraystretch{1.7}%
\begin{tabular}{||cccccc||}
\hline 
  $i$  & $\beta_{i}^{\dagger}$  & $s_{i}$  & $e_{i}$  & $\mu_{i}$  & $\nu_{i}$ \tabularnewline
\hline\hline
0  & $\mathbb{1}$  & 1  & 0  & 0  & $\frac{1}{\alpha}\!\left(\!\sqrt{2}\!+\!1\!\right)$ \tabularnewline
\hline 
1  & $\frac{1}{2}\!\left(\!b_{1}^{\dagger}\!-\!b_{2}^{\dagger}\!+\!b_{3}^{\dagger}\!-\!b_{4}^{\dagger}\!\right)$  & $\alpha$  & 0  & 0  & $\frac{1}{\alpha}\!\sqrt{2}$ \tabularnewline
\hline 
2  & $\frac{1}{\sqrt{2}}\!\left(\!b_{1}^{\dagger}\!-\!b_{3}^{\dagger}\!\right)$  & $\alpha$  & $2$  & $\frac{\alpha}{\sqrt{2}}$  & $\frac{\alpha}{2\sqrt{2}}\!+\!\frac{3}{2\alpha}\!\left(\!1\!+\!\frac{1}{\sqrt{2}}\!\right)$ \tabularnewline
\hline 
3  & $\frac{1}{\sqrt{2}}\!\left(\!b_{2}^{\dagger}\!-\!b_{4}^{\dagger}\!\right)$  & $\alpha$  & $2$  & $\frac{\alpha}{\sqrt{2}}$  & $\frac{\alpha}{2\sqrt{2}}\!+\!\frac{3}{2\alpha}\!\left(\!1\!+\!\frac{1}{\sqrt{2}}\!\right)$ \tabularnewline
\hline 
4  & $\frac{1}{2}\!\left(\!b_{1}^{\dagger}\!+\!b_{2}^{\dagger}\!+\!b_{3}^{\dagger}\!+\!b_{4}^{\dagger}\!\right)$  & $\alpha$  & 4  & $\alpha$  & $\alpha\!\frac{\sqrt{2}-1}{2}\!+\!\frac{1}{\alpha}\!\left(\!\sqrt{2}\!+\!\frac{1}{2}\!\right)$ \tabularnewline
\hline 
5  & $\frac{1}{2}\left(\!b_{1}^{\dagger}b_{3}^{\dagger}\!+\!b_{2}^{\dagger}b_{4}^{\dagger}\!\right)$  & 1  & $4\!-\!2\sqrt{2}$  & $\frac{1}{\alpha}\!\left(\!\sqrt{2}\!-\!1\!\right)$  & $\frac{1}{d\alpha}\!\left(\!\sqrt{2}\!-\!1\!\right)$\tabularnewline
 & $-\frac{1}{2\sqrt{2}}\!\left(\!b_{1}^{\dagger}b_{2}^{\dagger}\!+\!b_{2}^{\dagger}b_{3}^{\dagger}\!+\!b_{3}^{\dagger}b_{4}^{\dagger}\!+\!b_{1}^{\dagger}b_{4}^{\dagger}\!\right)$  &  &  &  & \tabularnewline
\hline 
6  & $\frac{1}{\sqrt{2}}\left(\!b_{1}^{\dagger}b_{3}^{\dagger}\!-\!b_{2}^{\dagger}b_{4}^{\dagger}\!\right)$  & $\alpha^{2}$  & 4  & $\alpha\!\left(\!1\!+\!\frac{1}{\sqrt{2}}\!\right)$  & $\frac{\alpha}{d}\!\left(\!1\!+\!\frac{1}{\sqrt{2}}\!\right)$ \tabularnewline
\hline 
7  & $\frac{1}{\sqrt{2}}\left(\!b_{1}^{\dagger}b_{2}^{\dagger}\!-\!b_{3}^{\dagger}b_{4}^{\dagger}\!\right)$  & $\alpha^{2}$  & 4  & $\alpha\!\left(\!1\!+\!\frac{1}{\sqrt{2}}\!\right)$  & $\frac{\alpha}{d}\!\left(\!1\!+\!\frac{1}{\sqrt{2}}\!\right)$ \tabularnewline
\hline 
8  & $\frac{1}{\sqrt{2}}\left(\!b_{1}^{\dagger}b_{4}^{\dagger}\!-\!b_{2}^{\dagger}b_{3}^{\dagger}\!\right)$  & $\alpha^{2}$  & 4  & $\alpha\!\left(\!1\!+\!\frac{1}{\sqrt{2}}\!\right)$  & $\frac{\alpha}{d}\!\left(\!1\!+\!\frac{1}{\sqrt{2}}\!\right)$ \tabularnewline
\hline 
9  & $\frac{1}{2}\!\left(\!b_{1}^{\dagger}b_{4}^{\dagger}\!+\!b_{2}^{\dagger}b_{3}^{\dagger}\!-\!\left(\!b_{1}^{\dagger}b_{2}^{\dagger}\!+\!b_{3}^{\dagger}b_{4}^{\dagger}\!\right)\!\right)$  & $\alpha^{2}$  & 4  & $\alpha\!\left(\!1\!+\!\frac{1}{\sqrt{2}}\!\right)$  & $\frac{\alpha}{d}\!\left(\!1\!+\!\frac{1}{\sqrt{2}}\!\right)$ \tabularnewline
\hline 
10  & $\frac{1}{2}\left(\!b_{1}^{\dagger}b_{3}^{\dagger}\!+\!b_{2}^{\dagger}b_{4}^{\dagger}\!\right)$  & $\alpha^{2}$  & $4\!+\!2\sqrt{2}$  & $\frac{1}{\alpha}\!\left(\!\sqrt{2}\!+\!1\!\right)$  & $\frac{1}{d\alpha}\!\left(\!\sqrt{2}\!+\!1\!\right)$\tabularnewline
 & $+\frac{1}{2\sqrt{2}}\!\left(\!b_{1}^{\dagger}b_{2}^{\dagger}\!+\!b_{2}^{\dagger}b_{3}^{\dagger}\!+\!b_{3}^{\dagger}b_{4}^{\dagger}\!+\!b_{1}^{\dagger}b_{4}^{\dagger}\!\right)$  &  &  &  & \tabularnewline
\hline 
11  & $\frac{1}{2}\!\left(\!b_{2}^{\dagger}b_{3}^{\dagger}b_{4}^{\dagger}\!-\!b_{1}^{\dagger}b_{3}^{\dagger}b_{4}^{\dagger}\!\right.$  & $d\alpha$  & 4  & $\frac{d}{\alpha}\sqrt{2}$  & 0 \tabularnewline
 & $\left.+b_{1}^{\dagger}b_{2}^{\dagger}b_{4}^{\dagger}\!-\!b_{1}^{\dagger}b_{2}^{\dagger}b_{3}^{\dagger}\!\right)$  &  &  &  & \tabularnewline
\hline 
12  & $\frac{1}{\sqrt{2}}\!\left(\!b_{2}^{\dagger}b_{3}^{\dagger}b_{4}^{\dagger}\!-\!b_{1}^{\dagger}b_{2}^{\dagger}b_{4}^{\dagger}\!\right)$  & $d\alpha$  & 6  & $\frac{d\alpha}{2\sqrt{2}}\!+\!\frac{3d}{2\alpha}\!\left(\!1\!+\!\frac{1}{\sqrt{2}}\!\right)$  & $\frac{d}{\alpha\sqrt{2}}$ \tabularnewline
\hline 
13  & $\frac{1}{\sqrt{2}}\!\left(\!b_{1}^{\dagger}b_{3}^{\dagger}b_{4}^{\dagger}\!-\!b_{1}^{\dagger}b_{2}^{\dagger}b_{3}^{\dagger}\!\right)$  & $d\alpha$  & 6  & $\frac{d\alpha}{2\sqrt{2}}\!+\!\frac{3d}{2\alpha}\!\left(\!1\!+\!\frac{1}{\sqrt{2}}\!\right)$  & $\frac{d}{\alpha\sqrt{2}}$ \tabularnewline
\hline 
14  & $\frac{1}{2}\!\left(\!b_{2}^{\dagger}b_{3}^{\dagger}b_{4}^{\dagger}\!+\!b_{1}^{\dagger}b_{3}^{\dagger}b_{4}^{\dagger}\!\right.$  & $d\alpha$  & 8  & $d\alpha\!\frac{\sqrt{2}-1}{2}\!+\!\frac{d}{\alpha}\!\left(\!\sqrt{2}\!+\!\frac{1}{2}\!\right)$  & $\frac{d}{\alpha}$ \tabularnewline
 & $\left.+b_{1}^{\dagger}b_{2}^{\dagger}b_{4}^{\dagger}\!+\!b_{1}^{\dagger}b_{2}^{\dagger}b_{3}^{\dagger}\!\right)$  &  &  &  & \tabularnewline
\hline 
15  & $b_{1}^{\dagger}b_{2}^{\dagger}b_{3}^{\dagger}b_{4}^{\dagger}$  & $\alpha^{2}$  & 8  & $\frac{\alpha}{d}\left(1+\sqrt{2}\right)$ & 0 \tabularnewline
\hline
\end{tabular}
\caption{Listing of all relevant properties of the one cycle eigenstates $\beta_{i}^{\dagger}\vert0\rangle$}
\label{tableHCB} 
\end{table}

To be able to evaluate the second term, we first specify our choice
of $S$: 
\begin{align}
S_{kk}=\prod_{C}s_{k}(C),\label{weighting}
\end{align}
where $s_{k}(C)=s_{i}$ only depends on the one cycle state $i$ which
occupies $C$ in $\chi_{k}$. Besides the restriction $s_{i}\neq0$
for all $i,$ the $s_{i}$ can be chosen freely and our choice is
listed in table (\ref{tableHCB}). We then obtain 
\begin{align}
\begin{split} & \sum_{i\neq k}\vert S_{ii}H_{1;ik}S_{kk}^{-1}\vert\\
= & \Vert S\left(H_{1}-H_{kk}\delta_{kk}\right)S^{-1}\chi_{k}\Vert_{1}\\
= & \sum_{\langle C,C'\rangle}\Vert SP_{1}H_{C,C'}S^{-1}\chi_{k}\Vert_{1}\\
\le & \sum_{\langle C,C'\rangle}\Vert SH_{C,C'}S^{-1}\chi_{k}\Vert_{1}\\
= & \sum_{\langle C,C'\rangle}t'\Vert S\left(p_{C'}^{\dagger}(C)p_{C}(C')+p_{C}^{\dagger}(C')p_{C'}(C)\right)S^{-1}\chi_{k}\Vert_{1}\\
= & \sum_{\langle C,C'\rangle}t'\left(\Vert Sp_{C'}^{\dagger}(C)p_{C}(C')S^{-1}\chi_{k}\Vert_{1}+\Vert Sp_{C}^{\dagger}(C')p_{C'}(C)S^{-1}\chi_{k}\Vert_{1}\right).\label{offDiag}
\end{split}
\end{align}
Looking at one of these terms specifically: 
\begin{align}
\begin{split} & \;\;\;\;\:\Vert Sp_{C'}^{\dagger}(C)p_{C}(C')S^{-1}\chi_{k}\Vert_{1}\\
 & =\Vert S'p_{C'}^{\dagger}(C)S'^{-1}\beta_{k}^{\dagger}(C)\vert0\rangle\Vert_{1;C}\Vert S'p_{C}(C')S'^{-1}\beta_{k}^{\dagger}(C')\vert0\rangle\Vert_{1;C'}\\
 & =:\nu_{k}(C)\mu_{k}(C').
\end{split}
\end{align}
Here $S'$ is the restriction of $S$ on single cycles: $S'_{ij}=s_{i}\delta_{ij}$.
The defined $\nu$ does not depend on $C'$ since $C'$ only determines
the direction from which the particles are hopping to $C$ but $\nu$
is invariant under discrete rotations (as can be seen in table \ref{tableHCB}).
As $\nu_{k}(C)$ only depends on the one cycle state $i$, which occupies
$C$ in $\chi_{k},$ we can once again write $\nu_{k}(C)=\nu_{i}$.
Accordingly for $C'$ occupied by the $j$th one cycle state we write
$\mu_{k}(C)=\mu_{j}$. All values of $\mu$ and $\nu$ are listed
in table \ref{tableHCB}. (\ref{offDiag}) is then bounded by 
\begin{equation}
\sum_{i\neq k}\vert S_{ii}H_{1;ik}S_{kk}^{-1}\vert\le\sum_{\langle C,C'\rangle}\left(\mu_{k}(C)\nu_{k}(C')+\mu_{k}(C')\nu_{k}(C')\right)t'
\end{equation}
and together with (\ref{Diag}) 
\begin{align}
\begin{split} & \;\;\;\;\:\vert H_{1;kk}-\lambda\vert-\sum_{i\neq k}\vert S_{ii}H_{1;ik}S_{kk}^{-1}\vert\\
 & \ge\!-2Nt\!-\!\lambda\!+\!\sum_{C}e_{k}(C)t-\sum_{\langle C,C'\rangle}\left(\mu_{k}(C)\nu_{k}(C')+\mu_{k}(C')\nu_{k}(C')\right)t'\\
 & =\!-2Nt\!-\!\lambda\!+\!\sum_{\langle C,C'\rangle}\!\left(\frac{e_{k}(C)}{c(C)}+\frac{e_{k}(C')}{c(C')}\right)t-\left(\mu_{k}(C)\nu_{k}(C')+\mu_{k}(C')\nu_{k}(C)\right)t'\\
 & =:\!-2Nt\!-\!\lambda\!+\!\sum_{\langle C,C'\rangle}\Lambda_{k}\left(C,C'\right).
\end{split}
\end{align}
This expression has a great advantage: Let $C$($C'$) be occupied
by the $i$th($j$th) one cycle state in $\chi_{k}$, for $t'<\sqrt{\frac{2e_{5}}{a}}\frac{t}{c(G)},a=\frac{1425+974\sqrt{2}}{896}$
our free parameters can then be chosen such that 
\begin{align}
\begin{split} & \Lambda_{k}\left(C,C'\right)\\
= & \left(\frac{1}{c(C)}e_{k}(C)+\frac{1}{c(C')}e_{k}(C')\right)t-\left(\mu_{k}(C)\nu_{k}(C')+\mu_{k}(C')\nu_{k}(C)\right)t'\\
= & \left(\frac{1}{c(C)}e_{i}+\frac{1}{c(C')}e_{j}\right)t-\left(\mu_{i}\nu_{j}+\mu_{j}\nu_{i}\right)t'\\
\ge & \frac{1}{c(G)}\left(e_{i}+e_{j}\right)t-\left(\mu_{i}\nu_{j}+\mu_{j}\nu_{i}\right)t'\\
=: & \Lambda_{i,j}\\
\ge & 0
\end{split}
\end{align}
for all $i,j$ and $\Lambda_{i,j}>0$ for $i\neq0,1$. (We will explicitly
calculate one such set of parameters later on). Hence we can drop
arbitrary summands and by doing so only decrease the sum. Since $N>\vert F_{1}(G)\vert$
there is at least one non-empty cycle not occupied by the one particle
ground state. Let $C_{1}$ be a such a cycle in state $\chi_{k}$
and $t'<\sqrt{\frac{2e_{5}}{a}}\frac{t}{c(G)}$. First we look at
the case where $C_{1}$ is occupied by a state $i$ with either three
or four particles or with two particles, not in their ground state,
and let $u(C_{1})$ be the set of cycles neighbouring $C_{1}$. Then,
by definition, $u(C_{1})$ has $c(C_{1})$ elements and we can calculate
\begin{align}
\begin{split} & \;\;\sum_{\langle C,C'\rangle}\Lambda_{k}\left(C,C'\right)\\
\ge & \sum_{C'\in u(C_{1})}\Lambda_{k}\left(C_{1},C'\right)\\
= & e_{k}(C_{1})t+\sum_{C'\in u(C_{1})}\frac{1}{c(C')}e_{k}(C')t-\left(\mu_{k}(C)\nu_{k}(C')+\mu_{k}(C')\nu_{k}(C)\right)t'\\
\ge & e_{i}t+\sum_{C'\in u(C_{1})}\min_{j}\left(\frac{1}{c(G)}e_{j}t-\left(\mu_{i}\nu_{j}+\mu_{j}\nu_{i}\right)t'\right)\\
= & e_{i}t+c(C_{1})\min_{j}\left(\frac{1}{c(G)}e_{j}t-\left(\mu_{i}\nu_{j}+\mu_{j}\nu_{i}\right)t'\right)\\
\ge & e_{i}t+c(G)\min_{j}\left(\frac{1}{c(G)}e_{j}t-\left(\mu_{i}\nu_{j}+\mu_{j}\nu_{i}\right)t'\right)\\
= & c(G)\min_{j}\left(\frac{1}{c(G)}\left(e_{i}+e_{j}\right)t-\left(\mu_{i}\nu_{j}+\mu_{j}\nu_{i}\right)t'\right)\\
= & c(G)\min_{j}\left(\Lambda_{ij}\right).
\end{split}
\end{align}
The lower bound in the last inequality holds since $c(C_{1})\le c(G)$
and 
\begin{align}
\begin{split}\min_{j}\left(\frac{1}{c(G)}e_{j}t-\left(\mu_{i}\nu_{j}+\mu_{j}\nu_{i}\right)t'\right)\le & \left(\frac{1}{c(G)}e_{0}t-\left(\mu_{i}\nu_{0}+\mu_{0}\nu_{i}\right)t'\right)\\
= & -\mu_{i}\nu_{0}t'\\
\le & 0.
\end{split}
\end{align}
Now let $C_{1}$ be occupied by a one particle non ground state (still
called $i$). Then the calculation still applies but in addition there
needs to be at least one additional cycle $C_{2}$ with at least two
particles since otherwise the state would contain at maximum $N-1$
particles. W.l.o.g. $C_{2}$ can be assumed to be occupied by the
two particle ground state ($l=5$), otherwise the first calculation
applies. We need to consider both the case, where $C_{1}$ and $C_{2}$
are neighbouring each other and where they are not; we start with
the latter and obtain in the same way as in the last calculation:
\begin{align}
\sum_{\langle C,C'\rangle}\Lambda_{k}\left(C,C'\right) & \ge c(G)\left(\min_{j}\left(\Lambda_{ij}\right)+\min_{m}\left(\Lambda_{lm}\right)\right).\end{align}
For $C_{1}$ and $C_{2}$ neighbouring each other we obtain: 
\begin{align}
\begin{split} & \sum_{\langle C,C'\rangle}\Lambda_{k}\left(C,C'\right)\\
\geq & \sum_{C_{2}\neq C'\in n(C_{1})}\Lambda_{k}\left(C_{1},C'\right)+\sum_{C_{1}\neq C'\in n(C_{2})}\Lambda_{k}\left(C_{1},C'\right)+\Lambda_{k}\left(C_{1},C_{2}\right)\\
\geq & (c(G)-1)\left(\min_{j}\left(\Lambda_{ij}\right)+\min_{m}\left(\Lambda_{lm}\right)\right)+\Lambda_{il}.
\end{split}
\end{align}
For the remaining case let $C_{1}$ be occupied by the two particle
ground state, then there needs to be at least one additional cycle
$C_{2}$ not in the one particle ground state or otherwise $\chi_{k}$
would be in $\Omega_{0}$. W.l.o.g. $C_{2}$ can be assumed to be
occupied by the two particle ground state, too, or otherwise one of
the other calculations applies. We demand our parameters to be chosen
such that $\Lambda_{55}\ge2\min_{j}\left(\Lambda_{5}j\right)$ (which
essentially means $\nu_{5}\le\max\left(\nu_{0},\nu_{1}\right)$) Then,
regardless of the positions of $C_{1}$ and $C_{2}$ we obtain: 
\begin{align}
\sum_{\langle C,C'\rangle}\Lambda_{k}\left(C,C'\right)\ge2c(G)\min_{j}\left(\Lambda_{5j}\right)
\end{align}
Using all these lower boundaries we can now apply basic calculus to
see that for 
\begin{align}
\begin{split}\xi(t') & =\frac{1}{\alpha^{2}(t')}=\frac{a}{3+\sqrt{2}}\frac{t'}{t}c(G)\\
d & =4/3
\end{split}
\end{align}
$\lambda\le\left(-2(N-4)-4\sqrt{2}\right)t$ and $t'\le\sqrt{\frac{2e_{5}}{a}}\frac{t}{c(G)}$
(\ref{LB}) suffices 
\begin{align}
\begin{split} & \min_{1\le k\le n_{1}}\Big(\vert H_{1;kk}-\lambda\vert-\sum_{i\neq k}\vert S_{ii}H_{1;ik}S_{kk}^{-1}\vert\Big)\\
\ge & -2Nt-\lambda+2c(G)\Lambda_{5,0}\\
= & \left(-2N+2e_{5}\right)t-\lambda-\xi(t)\left(3+\sqrt{2}\right)c(G)t'\\
= & \left(-2N+2e_{5}\right)t-\lambda-a\frac{c^{2}(G)t'^{2}}{t}\label{H1}.
\end{split}
\end{align}
It should be noted that this $\xi$ is not the true minimal function
in $t'$ to obey all the constraints laid out before. It is however
the ideal choice if we demand $\xi$ to be linear in $t'$, which
captures the same qualitative behaviour as a general solution, but
is on the other hand both a lot easier to calculate and to handle
later on. To calculate $\Vert H_{10}\Vert$ we can once again make
use of the Hermitian invariance of the spectral norm and H itself
being Hermitian to arrive at 
\begin{align}
\begin{split}\Vert H_{10}\Vert & =\frac{1}{\eta^{2}}\Vert H_{01}\Vert\\ & =\frac{1}{\eta}b\sqrt{c(G)}t'.\label{H10}
\end{split}
\end{align}
Combining (\ref{H1}) and (\ref{H10}) then leads to: 
\begin{align}
\begin{split} & \inf\Lambda_{1}\\
\ge & \inf\Big\{\lambda\in\mathbb{R}\Big\vert\left(-2N+2e_{5}\right)t-\lambda-a\frac{c^{2}(G)t'^{2}}{t}\le\frac{1}{\eta}b\sqrt{c(G)}t'\Big\}\\
= & \left(-2N+2e_{5}\right)t-\left(\frac{1}{\eta}b\sqrt{c(G)}+a\frac{t'}{t}c^{2}(G)\right)t'\label{inf}
\end{split}
\end{align}
Now we can use (\ref{sup}) and (\ref{inf}) to finally arrive at
\begin{align}
\begin{split} & \sup\Lambda_{0}\\
= & \left(-2N+e_{5}\right)t+\eta b\sqrt{c(G)}t'\\
< & \left(-2N+2e_{5}\right)t-\left(\frac{1}{\eta}b\sqrt{c(G)}+a\frac{t'}{t}c^{2}(G)\right)t'\\
\le & \inf\Lambda_{1}.
\end{split}
\end{align}
which holds for 
\begin{equation}
\tilde{t}<\dfrac{e_{5}}{a\tilde{t}c^{2}(G)+\left(\eta+\frac{1}{\eta}\right)b\sqrt{c(G)}}.
\end{equation}
We have introduced $\tilde{t}=\frac{t'}{t}$ for reasons of readability.This upper boundary becomes maximal for $\eta=1$ and the its supremum
$\tilde{t}_{\sup}$ can be found by solving the equality case for
$t'$: 
\begin{align}
\begin{split}\tilde{t}_{\sup} & =\left(\sqrt{\frac{b^{2}}{a^{2}c(G)}+\frac{e_{5}}{a}}-\frac{b}{a\sqrt{c(G)}}\right)\frac{1}{c(G)}\\
 & =0.6120\left(\sqrt{\frac{0.08478}{c(G)}+1}-\frac{0.2912}{\sqrt{c(G)}}\right)\frac{1}{c(G)}.
\end{split}
\end{align}
For a given $\tilde{t}<\tilde{t}_{\sup}$ we call $\eta_{-}(t')$
the infimum of all $\eta$ such that the upper bound of $\sup\Lambda_{0}$
is smaller than the lower bound of $\inf\Lambda_{1}$. It is given
by 
\begin{align}
\begin{split}\eta_{-}(t')= & \dfrac{2b\sqrt{c(G)}}{\left(e_{5}-a\tilde{t}^{2}c^{2}(G)\right)+\sqrt{\left(e_{5}-a\tilde{t}^{2}c^{2}(G)\right)^{2}-4b^{2}\tilde{t}^{2}c(G)}}\tilde{t}\\
= & \dfrac{b\sqrt{c(G)}}{e_{5}}\tilde{t}+\mathcal{O}\left(\tilde{t}^{3}\right).
\end{split}
\label{minimise}
\end{align}
Since $\Lambda_{0}$ and $\Lambda_{1}$ are separated we can apply
theorem 2 of \cite{MielkeHCB} with our norms. Therefore, we know
that for all eigenstates $\varphi_{k}$, $1\le k\le n_{0}$ of $H$
with eigenvalues in $\Lambda_{0}$ and $\eta>\eta_{-}(t')$: 
\begin{equation}
\eta\Vert P_{0}\varphi_{k}\Vert_{2}=\Vert P_{0}\varphi_{k}\Vert_{\Omega_{0}}>\Vert P_{1}\varphi_{k}\Vert_{\Omega_{1}}=\Vert P_{1}\varphi_{k}\Vert_{2}.
\end{equation}
Using $\Vert\varphi\Vert_{2}^{2}=\Vert P_{0}\varphi\Vert_{2}^{2}+\Vert P_{1}\varphi\Vert_{2}^{2}$
then achieves the third result: 
\begin{equation}
\Vert P_{0}\varphi_{k}\Vert_{2}\ge\frac{1}{\sqrt{1+\eta_{-}^{2}(t')}}\Vert\varphi_{k}\Vert_{2}.
\end{equation}
Since the lowest energies are in $\Lambda_{0}$ for $t'<t'_{\sup}$
we also obtained boundaries for them. These bounds are however by
no means optimal and can be improved with the help of the following
arguments.

Since any eigenvalue needs to be a member of either $\Lambda_{0}$
or $\Lambda_{1}$ all eigenvalues suffice 
\begin{align}
E_{k}\geq\min\left\{ \inf\Lambda_{0},\inf\Lambda_{1}\right\} .
\end{align}
This holds true for all $\eta>0$ regardless of the sets overlapping
or not. For a given $t'$ the lower boundary found this way becomes
maximal for 
\begin{align}
\eta=\eta_{+}(t')=\dfrac{2b\sqrt{c(G)}}{\left(e_{5}-a\tilde{t}^{2}c^{2}(G)\right)+\sqrt{\left(e_{5}-a\tilde{t}^{2}c^{2}(G)\right)^{2}+4b^{2}\tilde{t}^{2}c(G)}}\tilde{t}.
\end{align}
Hence a lower bound for all energy eigenvalues is given by 
\begin{align}
E_{k}\geq-\left(2N-e_{5}\right)t-b\sqrt{c(G)}\eta_{+}(t')t'.
\end{align}
To find an upper boundary on the energy of the $\vert F_{1}(G)\vert$
lowest states, we first observe that since $H_{0}$ is diagonal for
any normalised $\psi\in\Omega_{0}$ its expectation value under $H$
is given by 
\begin{align}
\left\langle \psi\vert H\vert\psi\right\rangle =-\left(2N-e_{5}\right)t.\label{expvalue}
\end{align}
We now choose a orthonormal base $\{\varphi_{k}\}$ of eigenstates
of $H,$ such that their energies are monotonously increasing, i.e.
$E_{k}\le E_{k+1}$ for all $k$. Then we rewrite the (normalised)
elements of $\Omega_{0}$ in terms of this basis: 
\begin{align}
\psi=\sum_{k}a_{k}\varphi_{k},
\end{align}
with $\sum_{k}\vert a_{k}\vert^{2}=1.$ Since $\Omega_{0}$ is $\vert F_{1}(G)\vert$
dimensional, it is possible to find a (again normalised) $\bar{\psi}\in\Omega_{0}$
with coefficients $\bar{a}_{k}$ such that its first $\vert F_{1}(G)\vert-1$
ones are all vanishing. Combining this with (\ref{expvalue}) leads
to 
\begin{align}
\begin{split}-\left(2N-e_{5}\right)t & =\left\langle \psi\vert H\vert\bar{\psi}\right\rangle \\
 & =\sum_{k=\vert F_{1}(G)\vert}\vert\bar{a}_{k}\vert^{2}E_{k}\\
 & \ge E_{\vert F_{1}(G)\vert}.
\end{split}
\end{align}
And as the $E_{k}$ are chosen to increase monotonously, the inequality
holds true for all $k\le\vert F_{1}(G)\vert$. Overall the bound for
the energies of the $\vert F_{1}(G)\vert$ lowest states then reads
\begin{align}
-\left(2N-e_{5}\right)t-b\sqrt{c(G)}\eta_{+}(t')t'\le E_{k}\le-\left(2N-e_{5}\right)t,
\end{align}
hereby having proven the first result. On the other hand, we can also
maximise $\eta$ in the same manner as in (\ref{minimise}) to find
an optimised lower boundary for the energy of the eigenstates in $\Lambda_{1}$.
For a fixed $t'$ it is given by $\sup\eta=\eta_{-}^{-1}(t')$ and
we obtain an energy gap between the lowest states and the upper states
of at least 
\begin{align}
\begin{split}\Delta\geq & e_{5}t-\left(a\tilde{t}c^{2}(G)+b\sqrt{c(G)}\eta_{-}(t')\right)t'\\
= & e_{5}t+\mathcal{O}\left(\tilde{t}^{2}\right),
\end{split}
\end{align}
which concludes the proof.

\subsection{Implication on graphs possessing a local reflection symmetry}From the proposition in section 2.1 we have seen that the ground state
of the chequerboard chain and the lattice of intertwined rhombi chains
with (D)PBC is at least $\vert F_{1}(G)\vert$ fold degenerated if
there is some overlap with the ground states of the decoupled system.
The main result states that under certain conditions, such an overlap
exists and that there are exactly $\vert F_{1}(G)\vert$ eigenvalues
in the lowest interval, which are separated from the rest of the spectrum.
Combining these two results, we obtain the following corollary:

\textbf{Corollary} \textit{For $t'<t'_{\sup}$ the chequerboard chain
with PBC, the line graph of the symmetric LIR (4,1) with DPBC and
the line graph of the symmetric LIR (3,2) with DPBC with $N$ hard
core bosons on $N-1$ cycles $C_{1},...,C_{N-1}$ have an exactly
$N-1$ fold degenerated ground state energy $E_{0},$ sufficing 
\begin{equation}
\left(-2(N-2)-2\sqrt{2}\right)t-b\sqrt{c(G)}\eta_{+}(t')t'\le E_{0}\le\left(-2(N-2)-2\sqrt{2}\right)t.
\end{equation}
The ground state is spanned by orthonormal states $\varphi_{k},$
sufficing $P_{0}\varphi_{k}=r(t')\varphi_{C_{k}}$ with $r(t')\ge\frac{1}{\sqrt{1+\eta_{-}^{2}(t')}}$.
Here $\varphi_{C_{k}}$ once again denotes the orthonormal base states
of $\Omega_{0}$ with two particles on cycle $C_{k}$ and one particle
on every other cycle. All other quantities are defined as in the main
theorem for $c(G)=2,$ $c(G)=6$ and $c(G)=3,$ respectively.}

\textit{Remarks:} We thereby have proven that the ground states of
the bosonic chequerboard chain and the line graphs of the LIRs are
dominated by localised pairs and that their band is indeed flat. Therefore,
we have found both one dimensional and two dimensional regular graphs
with rigorously provable localised repulsive bosonic pair formation.
To our knowledge this had not been achieved yet for either dimension.

As noted in chapter 2.1, we can generalise the statement $P_{0}\varphi_{k}=r(t')\varphi_{C_{k}}$
with $r(t')\ge\frac{1}{\sqrt{1+\eta^{2}(t')}}$ for the $\vert F_{1}(G)\vert$
lowest lying states to all graphs obeying the local symmetry for all
cycles. It should be noted that the lower dimension of the ground
states is not a contradiction to \cite{homogeneous}, as their result
only applies to the homogeneous case (See also chapter \ref{lowdensity}).\section{Potential improvements, generalisations and open questions}

\subsection{General model\label{GM}}

While the results of this paper already cover a broad class of line
graphs, extending the results to an even broader class of graphs would
be very desirable. Lifting or weakening the restrictions on cycles
lengths or interstitials would be one way of accomplishing this. As
already noted earlier, the longer cycles should not be a principal
problem, while the interstitials could indeed lead to some technical
as well as physical problems. One particular achievement would be
to solve both of these issues at the same time and include the kagome
chain or kagome lattice in the class of applicable graphs. Another
way would be to generalise the results to three dimensions. Especially
the chequerboard lattice generalises relatively straightforwardly
into a three dimensional analogous with cubes with hopping strength
$t$ on them as basic units and complete graphs with hopping strength
$t'$ between neighbouring faces. This model has compact eigenstates
on every cube with eigenvalue $-3t$ or even $-4t$ if one allows
for edges on the diagonals of the cube (but not on the diagonals of
its surfaces,) while the corresponding two particle ground states
on the cube have energies $-4\sqrt{2}t$ and $-\frac{12}{\sqrt{3}}t,$
respectively. This looks like a very encouraging start to any potential
future investigation.

Moving in another direction, one could also discuss additional terms
in the Hamiltonian. One interesting extension could be to allow for
repulsive next neighbour interactions defined in analogy to the hopping
strengths, with $V\ge0$ being the interaction on the cycles and $V'\ge0$
the interaction between the cycles. While the $V'=0$ case seems relatively
straightforward as it would only alter the one cycle ground states,
the $V'>0$ case would alter the model a lot more fundamentally, as
putting two particles into neighbouring cycles would in general make
them interact with each other and would therefore lower the critical
density by a varying amount depending on the geometry of the graph
in question.

\subsection{Fermionic models}

In addition to generalisations to the bosonic models it looks also
promising to investigate a potential translation of our result to
fermionic models. At first glance it seems most straight forward to
look into a spinless model, as its properties in regard to the occupationof single cycles are similar to the bosonic case. However, there are
multiple issues with this choice: The two particle ground state on
single cycles is not unique, for $t'\neq0$ there are non-vanishing
matrix elements between them and further issues arise. Adding the
next neighbour interactions discussed in \ref{GM} might be one solution
to these issues, but if one wants to consider a similar approach to
the one taken in this paper, a model with spin might be more promising.
In the hardcore limit the spin $\frac{1}{2}$ case has, in a certain
sense, a lot higher similarity with the bosonic case than the spinless
model. Unlike the spinless model, it has a unique singlet two particle
one cycle ground state following the same structure as the corresponding
ground state in the bosonic case. Therefore, a deeper analysis looks
very promising and might also lead to interesting discoveries on the
spin behaviour of the whole system. As our approach can in principle
be applied to any kind of particle, not allowing for a macroscopic occupation
of single cycles, one could even consider fermions with a finite (but
still repulsive) on site interaction. While one of course expects
the results to worsen as $U$ gets smaller, it should still be possible
for any given $U>0$ to prove a separation in the energy levels for
$t'$ sufficiently small.

\subsection{Hardcore bosons}

When looking at the details of the main theorem's proof, the reader
might have noticed that there are a few points where this work doesn't
completely maximise out the possibilities of the given approach. First
of all, any state with exactly two cycles with the two particle ground
state and all other particles in the one particle ground states will
have exactly one empty cycle and all remaining cycles will be occupied
by one particle ground states. Taking this into account complicates
the calculations of lower boundaries for $\sum_{\langle C,C'\rangle}\Lambda_{k}\left(C,C'\right)$
but could decrease the value of $a$ by around $\frac{1}{3}\frac{c(G)-1}{c(G)}$
and consequently slightly improve all dependent quantities.

A potentially bigger improvement can be achieved by allowing for more
free parameters in the definition of the $s_{i}$. Ideally, one would
want them to be chosen completely independent initially and would
allow them to be arbitrary functions in $t'$. However, it becomes
clear immediately that solving for that many free parameters with
the given constraints is not achievable analytically, hence a numerical
analysis would be very beneficial and might be able to improve the
applicability of the theorem well into the area of $t'\approx\frac{1}{c(G)}t$.

Nonetheless our work also shows some limits which the given approach
cannot exceed: Since $\Vert H_{01}\Vert$, $\Vert H_{10}\Vert$ and
$\Vert\left(H_{00}-\lambda I\right)^{-1}\Vert^{-1}$ are all calculated
exactly, even for the hypothetical (and unreachable) assumption that
$\Vert\left(H_{11}-\lambda I\right)^{-1}\Vert^{-1}$ does not depend
on $t'$ at all, one would still end up with $t'_{\sup}=\frac{e_{5}}{2b}t\approx\frac{1}{\sqrt{c(G)}}t$.
This upper limit cannot be breached by using the given norms and partition
of $H$. Also, the euclidean norm already tends to be a relatively
small matrix norm compared to other norms: For example using either
the $1$-norm or the $\infty$-norm with the same asymmetrical ansatz
only decreases the result because of $\Vert A\Vert_{2}\le\sqrt{\Vert A\Vert_{1}\Vert A\Vert_{\infty}}$
and the symmetry of the Hamiltonian, while choosing the 1-norm on
$\Omega_{0}$ and the $\infty$-norm on $\Omega_{1}$ (or the other
way around) completely breaks the result as either $\Vert H_{10}\Vert$
or $\Vert H_{10}\Vert$ becomes proportional to $N$. Choosing a more
fragmented partition of $H$ also tends to worsen the result since
having more non-diagonal matrices only leads to terms adding to the
boundaries of $\Vert\left(H_{ii}-\lambda I\right)^{-1}\Vert^{-1}$
and in many cases these boundaries can even become dependent on $N$
and thereby also breaking the result entirely. Hence, we conclude
that it seems very unlikely that any kind of Gershgorin-like argument
can reach the desired case of $t'_{\sup}\ge t$ even for $c(G)=2$
and presumably a completely new ansatz will have to be found for results
that include systems like the homogeneous chequerboard chain.

Beyond these incremental improvements, one can hope to prove the macroscopic
degeneracy of the ground state energy for more members of the class
of graphs, in which our main result holds and thereby concluding a
rigorous proof for the existence of localised pairs in the ground
state. Especially for the two dimensional chequerboard lattice with
DPBC such a result is highly suggested by the main result of this
paper in combination with its high symmetry; however, the lack of
a local symmetry as discussed for some of the graphs in our class
makes a formal proof much harder to achieve.

\end{document}